%% file: ms.tex
\documentclass[]{emulateapj}

\usepackage[utf8]{inputenc}
\usepackage{framed}
\input{commands.tex}

\begin{document}
\shorttitle{HD32963}
\shortauthors{Rowan \& Meschiari}

\title{The Lick-Carnegie Exoplanet Survey: HD~32963 -- A New Jupiter Analog Orbiting a Sun-like Star}
\author{Dominick Rowan \altaffilmark{1}, Stefano Meschiari \altaffilmark{2}, Gregory Laughlin\altaffilmark{3}, Steven S. Vogt\altaffilmark{3}, R. Paul Butler\altaffilmark{4},Jennifer Burt\altaffilmark{3}, Songhu Wang\altaffilmark{3,5}, Brad Holden\altaffilmark{3}, Russell Hanson\altaffilmark{3}, Pamela Arriagada\altaffilmark{4}, Sandy Keiser\altaffilmark{4}, Johanna Teske\altaffilmark{4}, Matias Diaz\altaffilmark{6}}
\altaffiltext{1}{
Byram Hills High School, Armonk, NY 10504
}
\altaffiltext{2}{
McDonald Observatory, University of Texas at Austin, Austin, TX 78712
}
 \altaffiltext{3}{UCO/Lick Observatory, Department of Astronomy and Astrophysics, University of California at Santa Cruz,Santa Cruz, CA 95064}
 \altaffiltext{4}{Department of Terrestrial Magnetism, Carnegie Institute of Washington, Washington, DC 20015}
 \altaffiltext{5}{School of Astronomy and Space Science and Key Laboratory of Modern Astronomy and Astrophysics in Ministry of Education, Nanjing University, Nanjing 210093, China}
 \altaffiltext{6}{Departamento de Astronomia, 
Universidad de Chile}
\email{stefano@astro.as.utexas.edu}

\begin{abstract}
We present a set of 109 new, high-precision Keck/HIRES radial velocity (RV) observations for the solar-type star HD~32963. Our dataset reveals a candidate planetary signal with a period of 6.49 $\pm$ 0.07 years and a corresponding minimum mass of 0.7 $\pm$ 0.03 Jupiter masses. Given Jupiter's crucial role in shaping the evolution of the early Solar System, we emphasize the importance of long-term radial velocity surveys. Finally, using our complete set of Keck radial velocities and correcting for the relative detectability of synthetic planetary candidates orbiting each of the 1,122 stars in our sample, we estimate the frequency of Jupiter analogs across our survey at approximately 3\%.
\end{abstract}


\input{introduction.tex}

\input{stellar.tex}

\input{fit.tex}

\input{frequency.tex}

\input{HD32963discussion.tex}

\acknowledgments
\input{acknowledgments.tex}

{\it Facilities:} \facility{Keck (HIRES)}

\bibliographystyle{apj}
\bibliography{biblio}

\end{document}

%% file: commands.tex
\usepackage{color}

\newcommand{\topp}{(\textit{Top})}
\newcommand{\bottomp}{(\textit{Bottom})}

\newcommand{\panp}[1]{(\textit{#1})}
\newcommand{\ms}{\ensuremath{\ \mathrm{m\ s}{^{-1}}}}
\newcommand{\mass}{\mathcal{M}}

\newcommand{\msun}{\ensuremath{\mass_\s\sun}}

\newcommand{\mjup}{\ensuremath{\mass_\s{J}}}

\newcommand{\Kepler}{\textit{Kepler}}
\newcommand{\s}[1]{\mathrm{#1}}

%% file: introduction.tex
\section{Introduction}\label{sec:intro}
The past decade has witnessed an astounding acceleration in the rate of exoplanet detections, thanks in large part to the advent of the \Kepler{} mission. As of this writing, more than 1,500 confirmed planetary candidates are known\footnote{See \url{http://stefano-meschiari.shinyapps.io/PlanetPlot} for an interactive mass/period/discovery method/discovery year plot.}. The excitement of the modern exoplanet age is largely driven by the discovery of low-mass, potentially-habitable planets detected through transit observations. 

\begin{figure}
\plotone{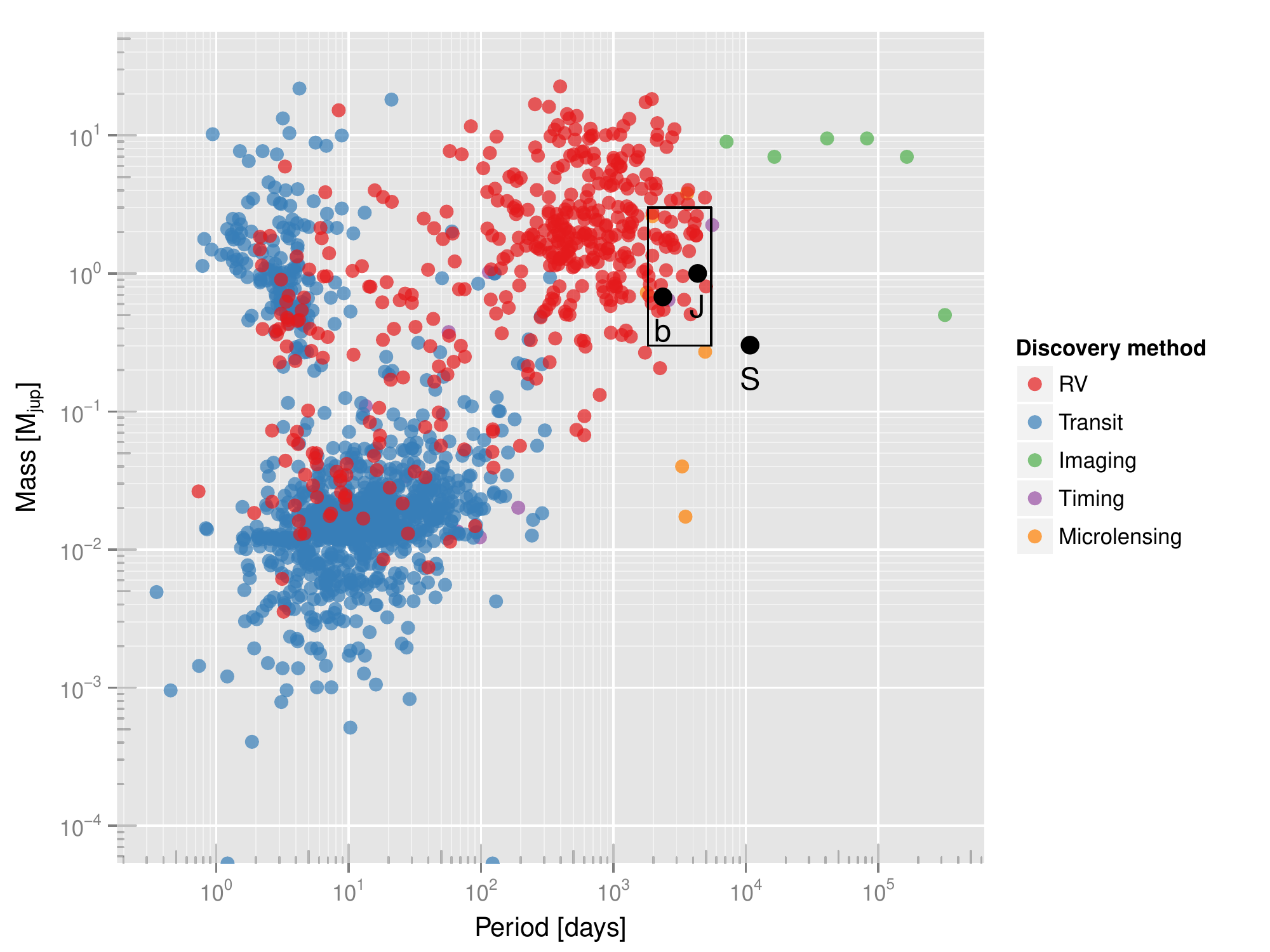}
\plotone{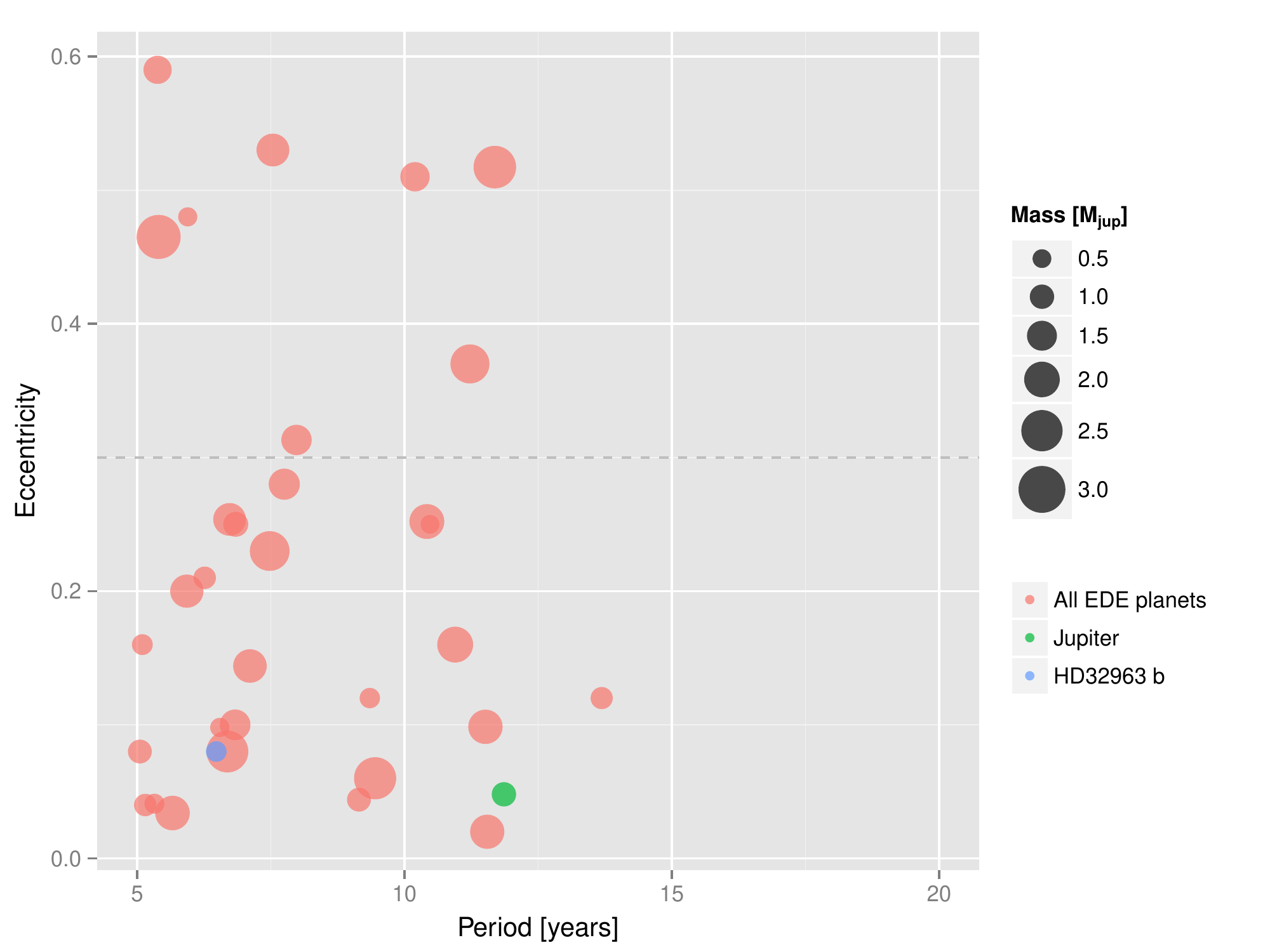}
\caption{(\textit{Top}) Period-mass diagram for the exoplanet candidates known as of August 2015 (data from the Exoplanetary Data Explorer). The box marks  the region of parameters that we use to define a Jupiter analog ($0.3 < \mass < 3\mjup$ and $5 < P < 20$ years. The labeled points represent Jupiter, Saturn, and HD~32963~b. (\textit{Bottom}) Period-eccentricity diagram of planets with $0.3 < \mass < 3\mjup$. We only consider planets with $e < 0.3$ as Jupiter analogs.}
\label{fig:population}
\end{figure}

Figure \ref{fig:population} shows a mass-period diagram of the known planetary candidates. As is well known, three broad but well-separated families of exoplanets that have no counterpart in the Solar System populate the mass-period diagram: (1) close-in super-Earths, (2) hot Jupiters with $P < 10$ days, and (3) exo-Jupiters with $P > 100$ days.

At first glance, it is tempting to conclude that the Solar System is an outlier within the galactic exoplanet census. \citet{Martin15}, however, argue that based on a number of metrics, the Solar System planets are relatively ``typical'', concluding that Solar System analogs (and, by extension, potentially life-bearing planetary systems) might be common after all. They note that the main properties that distinguish the Solar System from other exoplanetary systems are the absence of very close-in terrestrial planets at $P < 100$ days \citep[e.g.][]{Chiang13}, and the orbital location of Jupiter ($P = 11.86$  years, $a$ = 5.2 AU, $e$ = 0.05). They attribute the status of Jupiter as an outlier to observational bias, which favors short-period planets.

Given the importance of Jupiter in the dynamical history of our Solar System, the presence of a Jupiter analog is likely to be a crucial differentiator for exoplanetary systems. The dynamical evolution of Jupiter profoundly shaped the early history of the Solar System \citep[see, e.g. ][]{Morbidelli07}. Recently, \citet{Batygin15} suggested that if Jupiter migrated inwards early on, it would have increased the rate of collisions between planetesimals, initiating a collisional cascade \citep{kessler78}, and effectively clearing out the inner annulus of the Solar System \citep[as well as potentially explaining the low mass of Mars as well; ][]{Walsh11}. With the formation of Saturn, a mean motion resonance with Jupiter was created, reversing Jupiter's migration direction. In this picture, the Solar System was then left with Jupiter in a long-period orbit and small, gas-starved planetary bodies in the terrestrial region. The presence of a Jupiter-like planet in the system  may also influence the habitability of Earth analogs \citep[e.g.][]{Horner10}. As a consequence, substantiating whether Jupiter lies at the edge of the exo-Jupiter locus, or is a typical member of an as-yet undiscovered population, may be key to constraining the frequency of true Solar System analogs, and, by extension, worlds like Earth.

\begin{figure}
\plotone{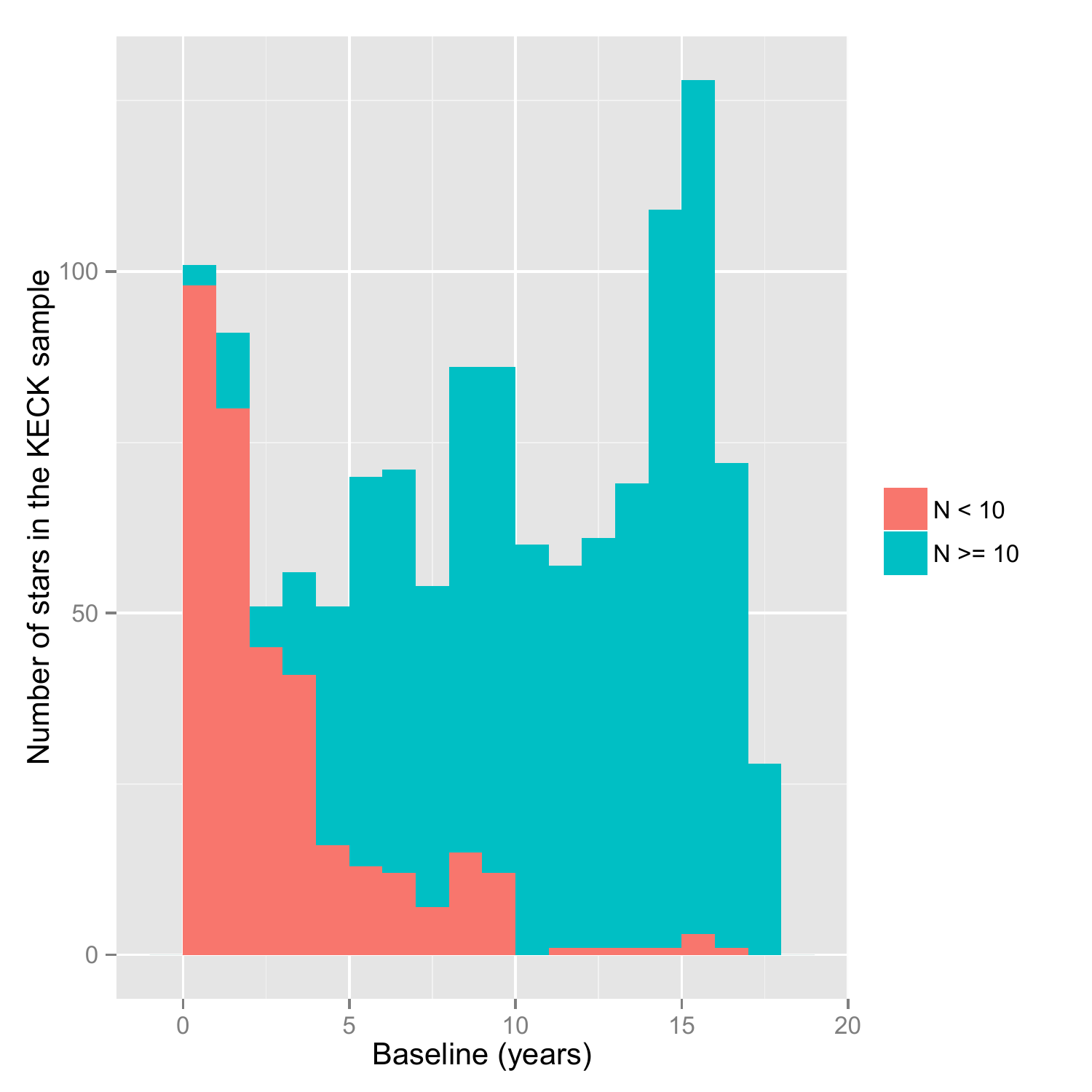}
\caption{Histogram of the RV baseline for stars in our Keck sample. We highlight stars that have more than 10 observations separately.}\label{fig:baseline}
\end{figure}

Currently, the most prolific technique to populate the exo-Jupiter region of the mass-period diagram  is radial velocity surveys. Unlike the transit method, which has a strong bias for short-period orbits, the radial velocity method can probe the outer regions of extraplanetary systems over the bulk of the parent star's lifetime. In order to safely detect a planetary signal, the minimum baseline is approximately 0.7 to 1 cycles \citep{Wittenmyer11}, which translates to $5 \lesssim T_\s{minimum} \lesssim 15$ years for Jupiter-like planets. This period range has become fully accessible by established RV surveys over the past few years. Figure \ref{fig:baseline} shows a histogram of the baseline of the RV observations of stars. In our Keck sample alone, there are 793 stars that have been observed for more than 5 years and 489 stars that have been observed more than 10 years. 


In this paper, we report that our set of 109 velocities obtained for the star HD~32963 (spanning approximately 16 years) strongly suggests the presence of a Jupiter analog with $P \approx 2372$ days (6.5 years), $\mass \approx 0.7$ \mjup, $K \approx 11\ \ms$, and $e \approx 0.07$. The host star can be considered a close solar analog \citep{PortodeMello14}, therefore making HD~32963 a benchmark for Jupiter analogs orbiting Sun-like stars.

Subsequently, we analyze a snapshot of our Keck sample of RV observations, and derive an occurrence rate of Jupiter analogs (defined here to lie within the parameter region with $5 \le P \le 15$ years, $0.3 \le \mass \le 3\mjup$, and $e \le 0.3$), corrected for the relative observability around each of the 1,122 stars in our sample. We find an occurrence rate of Jupiter analogs across our sample of approximately $3\%$ for the period range spanned by our observations, consistent with previous findings \citep{Wittenmyer11}. Under some reasonable assumptions about the underlying mass-period distribution of the Jupiter analogs, we find that the occurrence rate should lie between 1\% and 4\%. We conclude that the position of Jupiter on the ``fringe'' of the exo-Jupiter population is likely real, and that long-period Jupiter analogs are relatively rare.

The plan for the paper is as follows: in \S \ref{sec:parameters}, we review the stellar properties of HD~32963 and our Doppler technique. In \S \ref{sec:fit}, we discuss our radial velocity observations and the 1-planet model we use to interpret the RV variations exhibited by the star. In \S \ref{sec:frequency}, we present a simple calculation to derive an estimate for the occurrence rate of Jupiter analogs. In \S \ref{sec:discussion}, we discuss the HD~32963 system in light of the broader exoplanetary population and conclude.

%% file: stellar.tex
\section{Stellar parameters}\label{sec:parameters}

\begin{deluxetable}{rll}
\centering
\tablecaption{Stellar parameters for HD~32963\label{tab:params}}
\tablehead{{Parameter}&{Value}&{Reference}}
\tablecolumns{3}
\startdata
$T_\s{eff}$ ($K$) & 5727 $\pm$ 32 & \citep{Ramirez13} \\
$[Fe/H]$ & 0.11 $\pm$ 0.05 & \citep{Ramirez13} \\
log $g$ & 4.41 $\pm$ 0.03 &\citep{Ramirez13} \\
$S_\s{hk}$ & 0.159 $\pm$ 0.008 & This work \\
Age (Gyr) & 4.99$^{6.71}_{3.55}$ & \citep{Ramirez13} \\
Mass (\msun) & 1.03 & \citep{Ramirez13} \\
$V$ & 7.60 & \citep{Ramirez13} \\
$B-V$ & 0.6 & SIMBAD \\
$M_\s{bol}$ & 4.85 $\pm$ 0.09 & \citep{PortodeMello14} \\
\enddata
\end{deluxetable}

\begin{figure}
\centering
\plotone{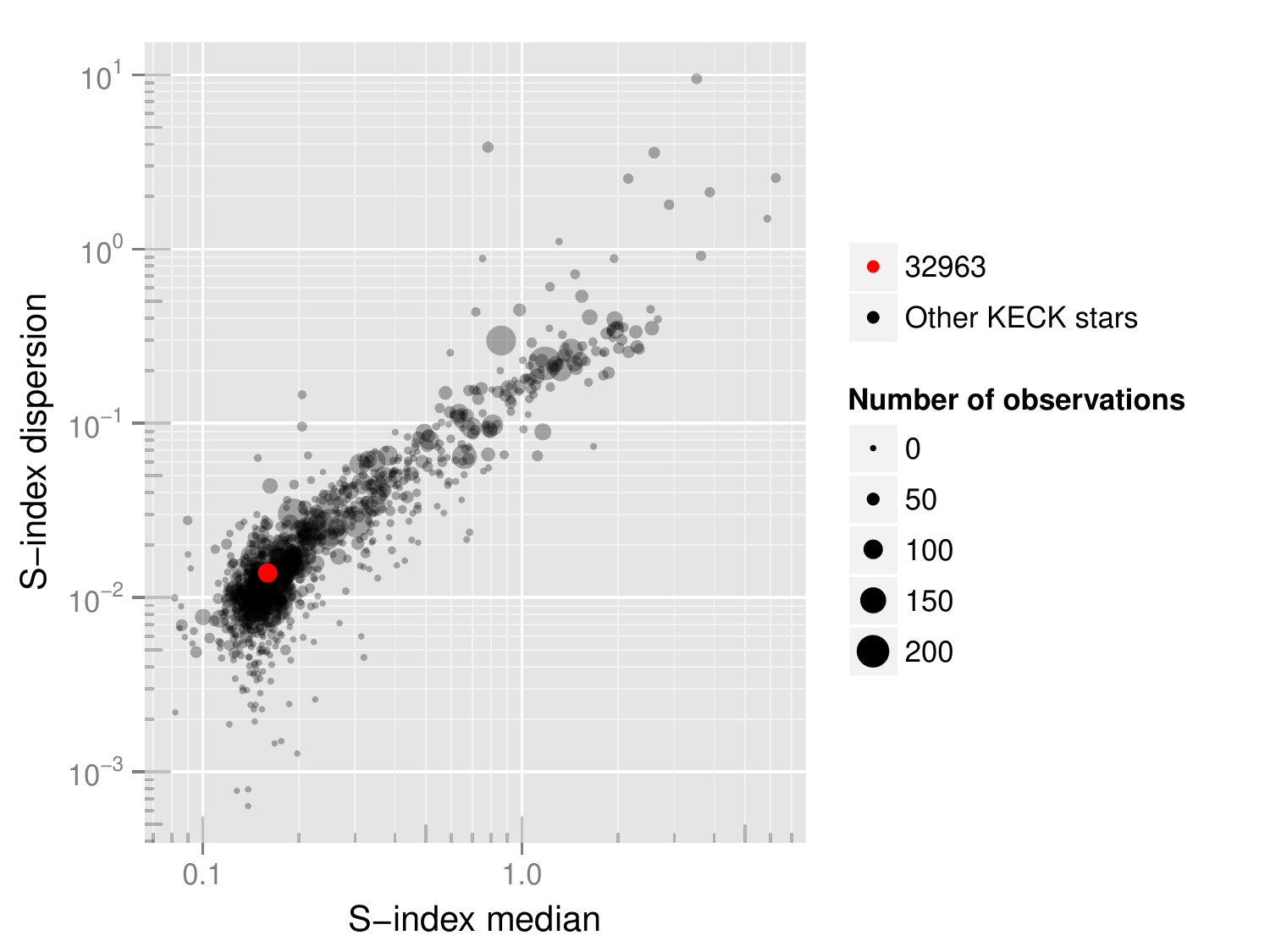}
\caption{The median $S$ value for the stars in the current Keck sample. HD~32963 is shown in red. The size of the points is proportional to the number of observations.}
\label{fig:sindex}
\end{figure}
\begin{figure}
\plotone{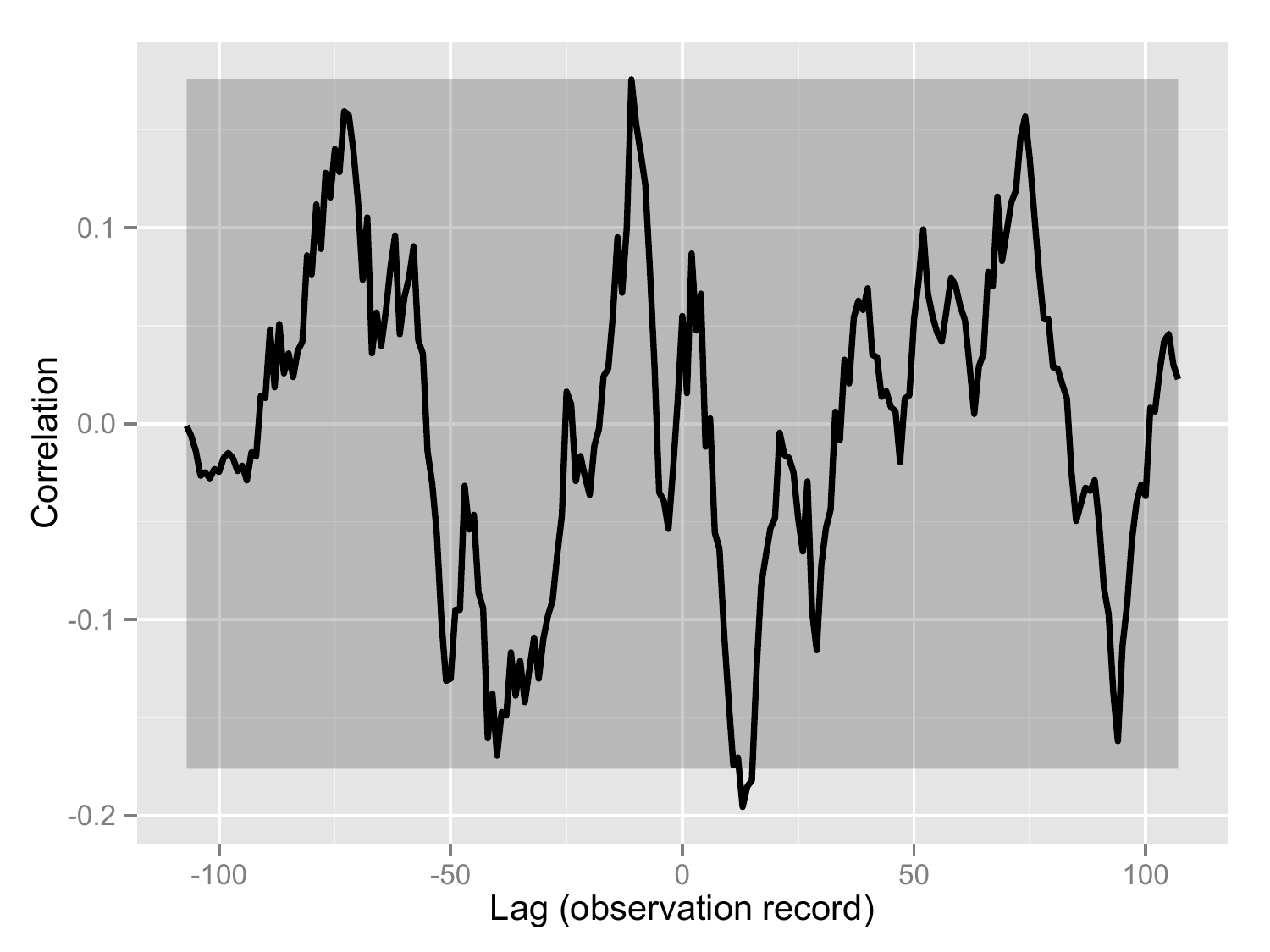}
\plotone{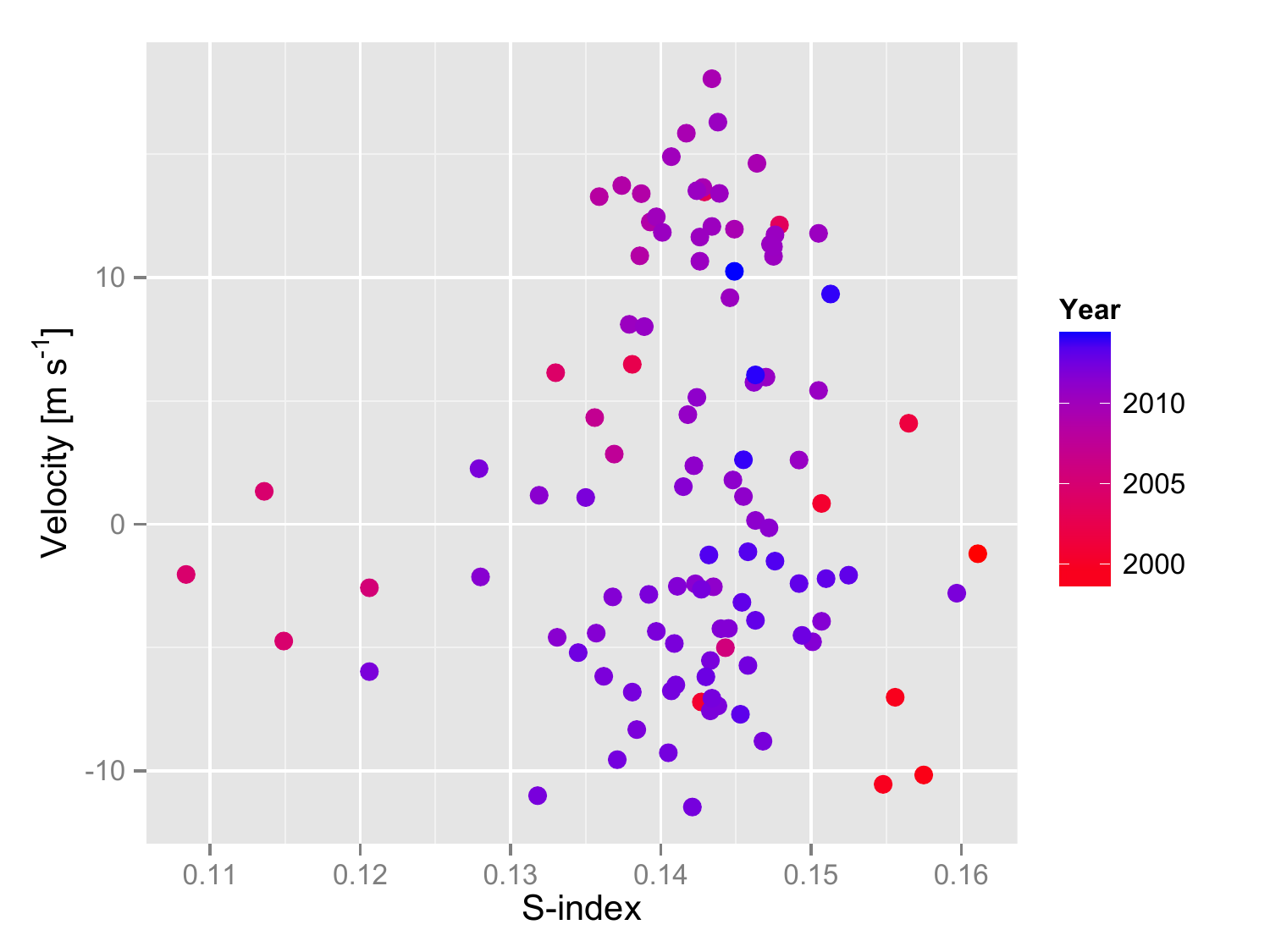}
\caption{\topp{} Correlation plot for RV data points and their associated S-index values. We compute the Pearson correlation cofficient between values, shifted by the specified lag in observation index. The shaded area marks the 95\% confidence interval for the Pearson correlation coefficient, estimated using sets of white noise data. \bottomp{} Scatter plot of the Keck radial velocity data and their associated $S$-index value. }\label{fig:scatter}
\end{figure}
\begin{figure}
\plotone{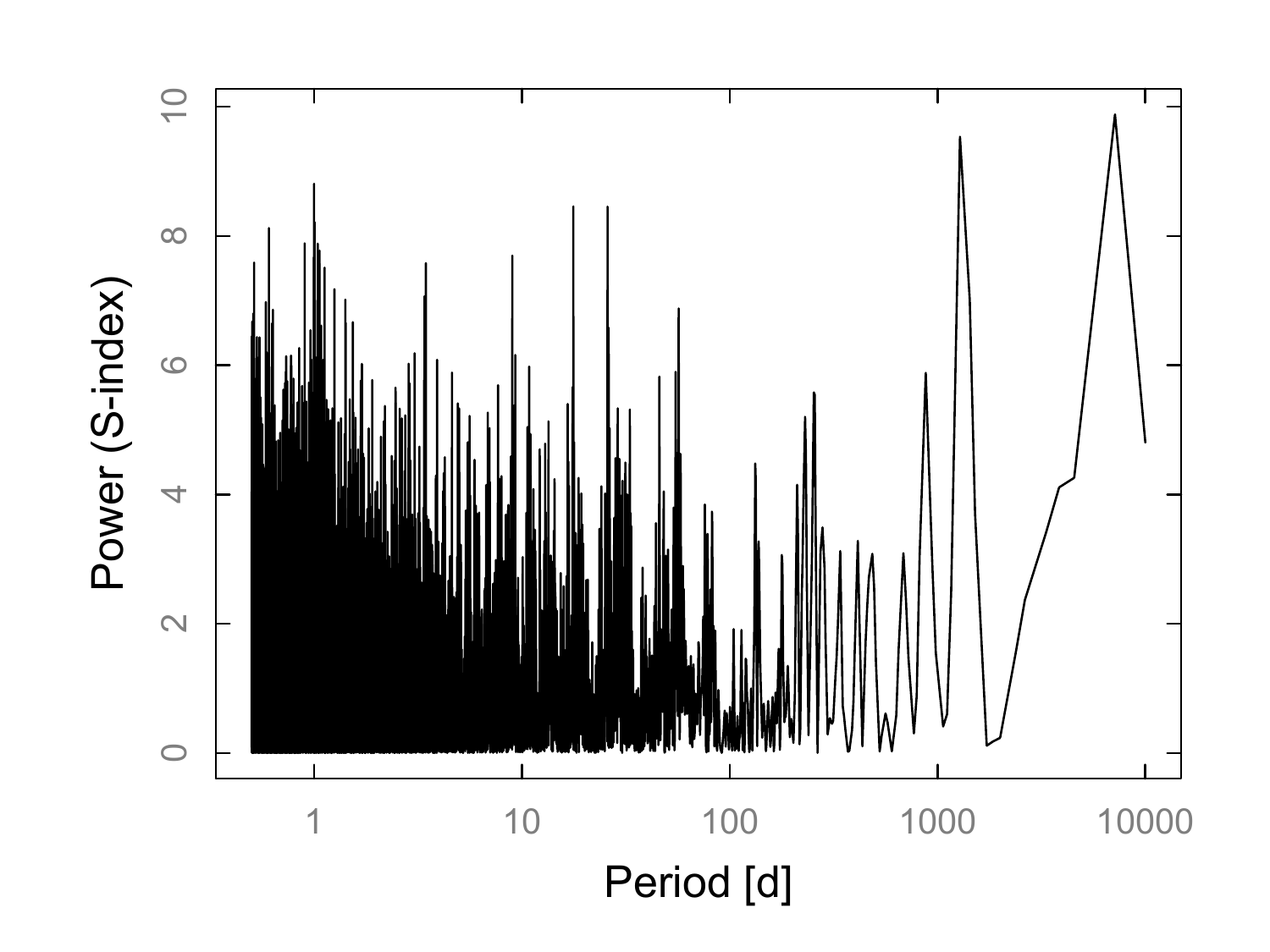}
\caption{Periodogram of the Mt. Wilson S-index values associated with our Keck spectra of HD 32963.}\label{fig:sindex_periodogram}
\end{figure}

HD 32963 (HIP 23884) is a bright (V = 7.6) star\footnote{We note that the spectral type of HD 32963 is reported as G5IV on the SIMBAD stellar database; however, the values of $T_\s{eff}$ and $\log g$ derived by \citet{Ramirez13} and \citet{PortodeMello14} are not consistent with this classification.}. Given \citet{PortodeMello14} included this star in their study of nearby stars exhibiting similarities in their stellar parameters to those of the Sun. Using photometric and spectroscopic observations, \citet{Ramirez13} derived $\mass \approx 1.03 \mass_\sun$,  effective temperature $T_\s{eff} \approx 5727$ K, and age $\approx$ 4.9 Gyr. Stellar parameters are reported in Table \ref{tab:params}.

In order to evaluate the level of stellar activity, we obtained Mt. Wilson $S$-index values for each radial velocity observation. The $S$-index measures the ratio of flux from 1${\rm \AA}$ bins surrounding the line centers of the Ca II H\& K lines (at 3968.47${\rm \AA}$ and 3933.66${\rm \AA}$), as compared to two broader 25${\rm \AA}$ bandpasses at 250${\rm \AA}$ to either side of the Ca II H\& K line location. The $S$-index is considered a proxy for chromospheric stellar activity, since it is correlated with the spot activity on the stellar surface. In turn, spots suppress radial velocity shifts due to convection, leading to a potential correlation between the $S$-index and RV measurements and a long-term Doppler signal. This Doppler signal may have amplitude and periodicity mimicking the Keplerian signal from a distant planet. Therefore, we check for correlations between the $S$-index and the RV measurements as a standard step to identify stars with an ongoing magnetic cycle \citep{Boisse2011}.

Figure \ref{fig:sindex} shows the median $S$ value for the stars in the Keck sample. \citet{Isaacson10} estimated a relatively low stellar jitter of approximately $2.615$\ms (in this case, jitter is defined as the difference in quadrature between the RMS of the velocities and the instrumental uncertainties). The top panel of Figure \ref{fig:scatter} shows a correlation plot of the radial velocity observations and their $S$-index values. The correlation is computed at different values of observational lag (a lag of 0 corresponds to the same epoch for each RV and S-index observation). There does not appear to be any significant degree of correlation. Finally, we note that the Lomb-Scargle periodogram of the $S$-index observations shown in Figure \ref{fig:sindex_periodogram} does not show any strong periodicity in the data (the strongest peak at 1,282 days has a false alarm probability of $\approx 14\%$). 

Based on the lack of strong periodicities in the $S$-index time series and of any significant correlation with the RV observations, we conclude it is unlikely that the $K\approx 10 \ms$ is due to stellar activity.

\subsection{Keck/HIRES RV observations}

\input{HD32963_vels}

We used the HIRES spectrometer \citep{Vogt94} of the Keck-I telescope for all the RV observations presented in this paper. Doppler shifts were obtained with the usual technique of impressing an iodine absorption spectrum on top of the collected light from the observed star \citep[see e.g.,][and others]{Vogt14, Burt14, Vogt15}.

Table \ref{tab:data} shows the complete set of our RV observations for HD~32963. Our dataset comprises 199 measurements over approximately 16 years (5838 days), binned into 109 measurements using 2-hour binning. The median formal uncertainty of the binned RV measurements is $\approx 1.2\ms$, with a velocity scatter of $\approx 8\ms$. Figure \ref{fig:data} presents the 109 individual Keck observations.

%% file: HD32963_vels.tex
\begin{table}
\centering
\caption{Radial Velocity observations (sample)} 
\begin{tabular}{cccc}
\hline
Time & Date & RV & Uncertainty \\ 
{[BJD]} &  & [$\ms$] & [$\ms$] \\
\hline
2450837.79 & 1998-01-24 & 0.31 & 1.27 \\ 
2451072.14 & 1998-09-15 & -8.66 & 1.50 \\ 
2451172.89 & 1998-12-25 & -5.51 & 1.36 \\ 
2451227.78 & 1999-02-18 & -9.04 & 1.54 \\ 
2451543.98 & 1999-12-31 & 2.35 & 1.34 \\ 
2451552.86 & 2000-01-09 & -5.69 & 1.41 \\ 
2451900.03 & 2000-12-21 & 5.60 & 1.22 \\ 
2452238.90 & 2001-11-25 & 7.99 & 1.36 \\ 
2452536.14 & 2002-09-18 & 14.98 & 1.56 \\ 
2452601.97 & 2002-11-23 & 13.64 & 1.55 \\ 
2452899.08 & 2003-09-16 & 7.66 & 1.59 \\ 
2453017.82 & 2004-01-13 & -0.53 & 1.60 \\ 
2453071.82 & 2004-03-07 & 2.84 & 1.45 \\ 
2453072.85 & 2004-03-08 & -3.23 & 2.23 \\ 
2453303.01 & 2004-10-24 & -0.82 & 1.35 \\ 
2453400.91 & 2005-01-30 & -3.59 & 1.21 \\ 
2453985.04 & 2006-09-06 & 5.73 & 1.12 \\ 
2454130.78 & 2007-01-30 & 4.35 & 1.28 \\ 
2454399.04 & 2007-10-25 & 15.01 & 1.50 \\ 
2454490.84 & 2008-01-25 & 12.48 & 0.99 \\ 
\hline
\end{tabular}
\label{tab:data}
\end{table}

%% file: fit.tex

\section{Best-fit Keplerian model}\label{sec:fit}
\begin{figure}
\centering
\plotone{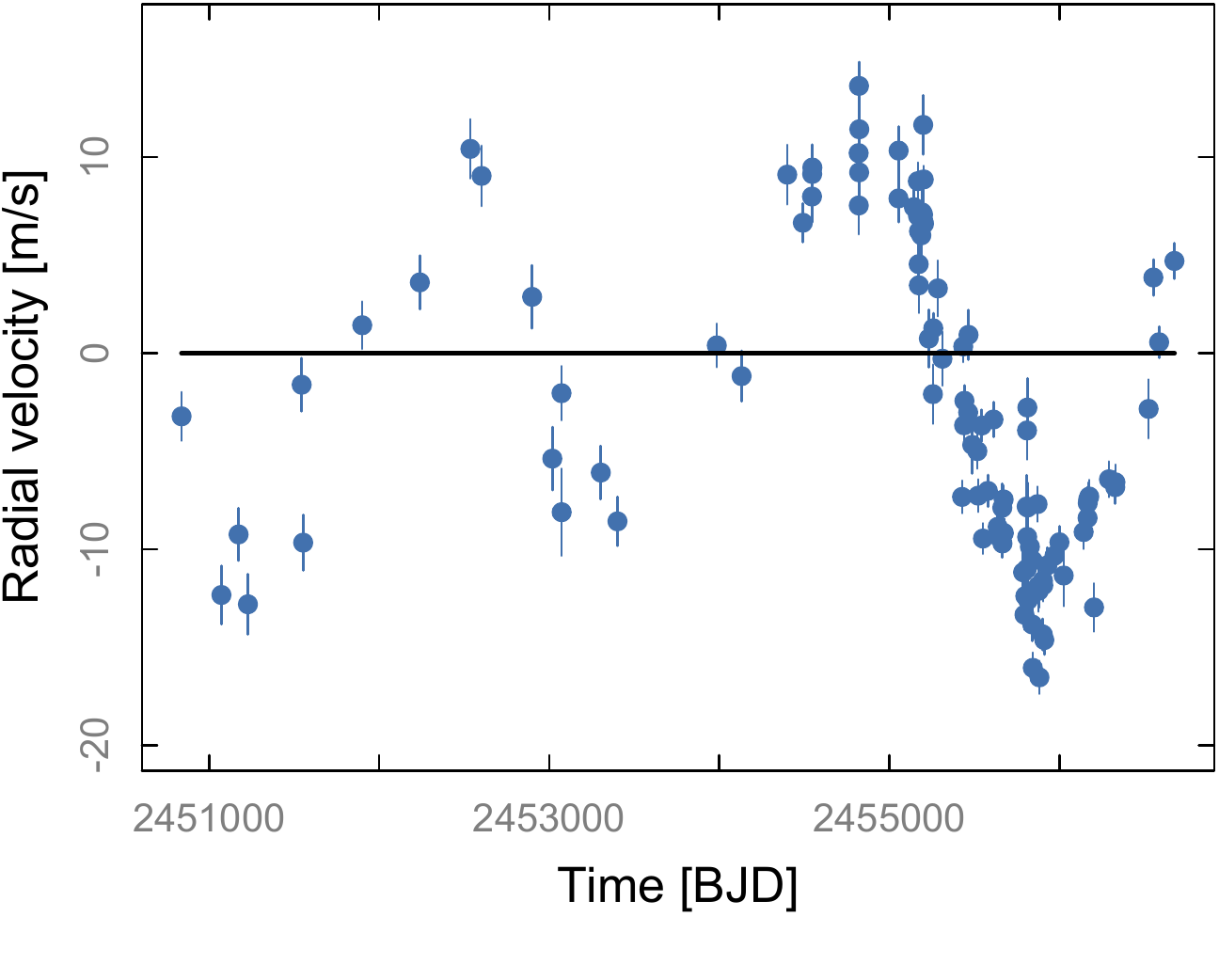}
\caption{\label{fig:data} Keck RV observations for HD~32963. The dataset spans approximately 16 years.}\
\end{figure}

\begin{figure}
\plotone{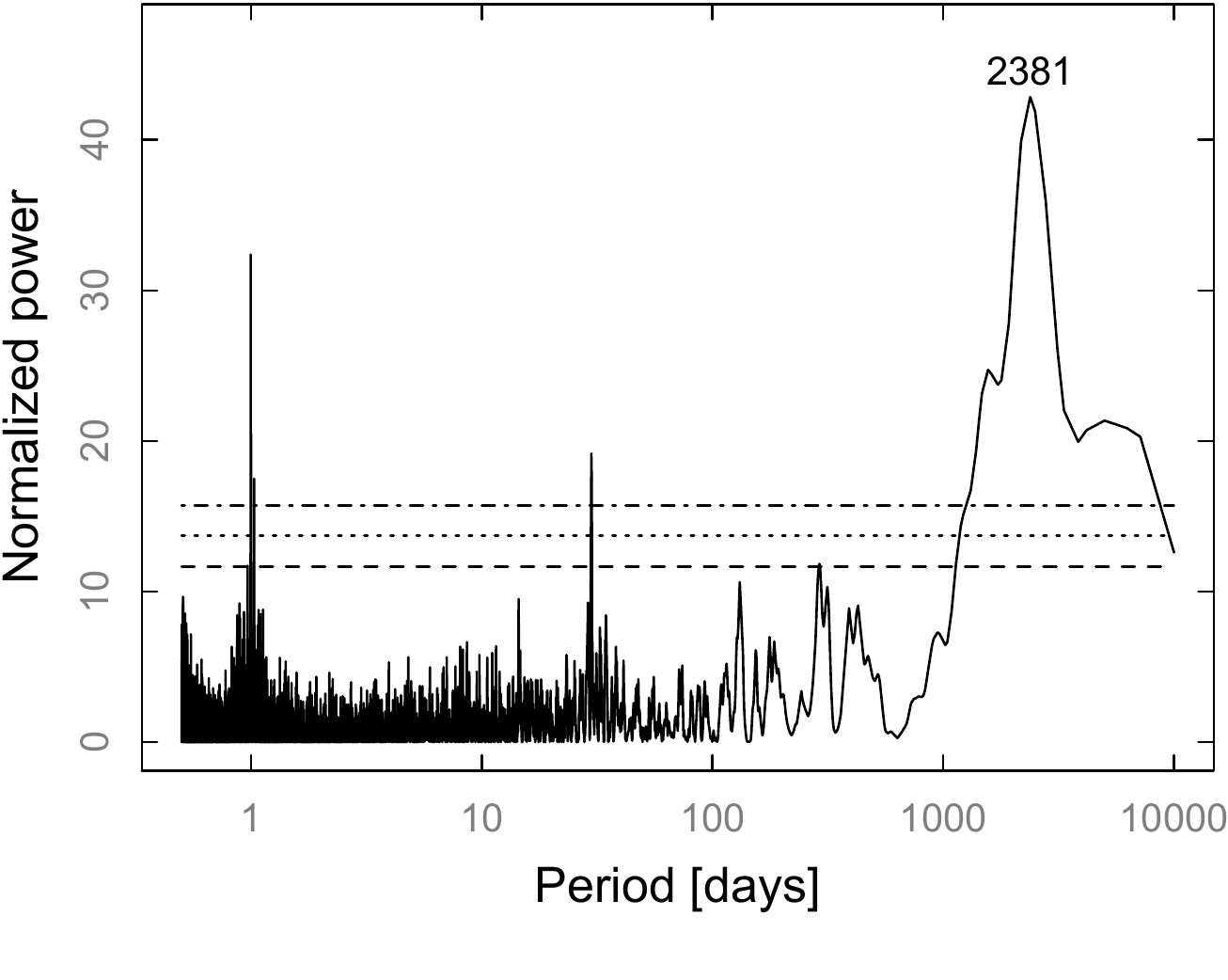}
\plotone{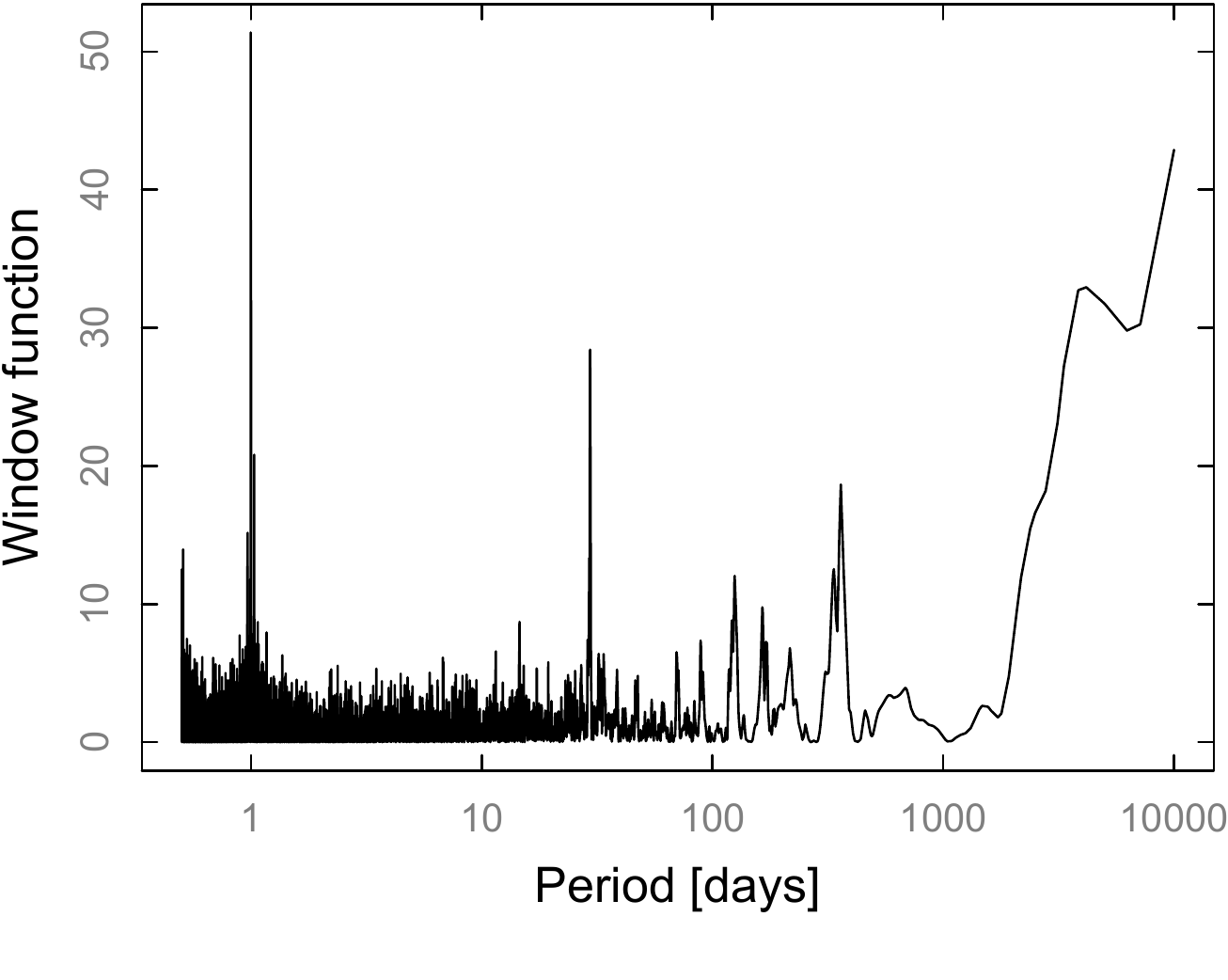}
\caption{\label{fig:periodograms}\topp{} Error-weighted Lomb-Scargle periodogram for HD 32963. False-alarm probability levels are shown at the 10\%, 1\% and 0.1\% level. \bottomp{} Spectral window function. }
\end{figure}

\begin{figure}
\centering
\epsscale{0.9}
\plotone{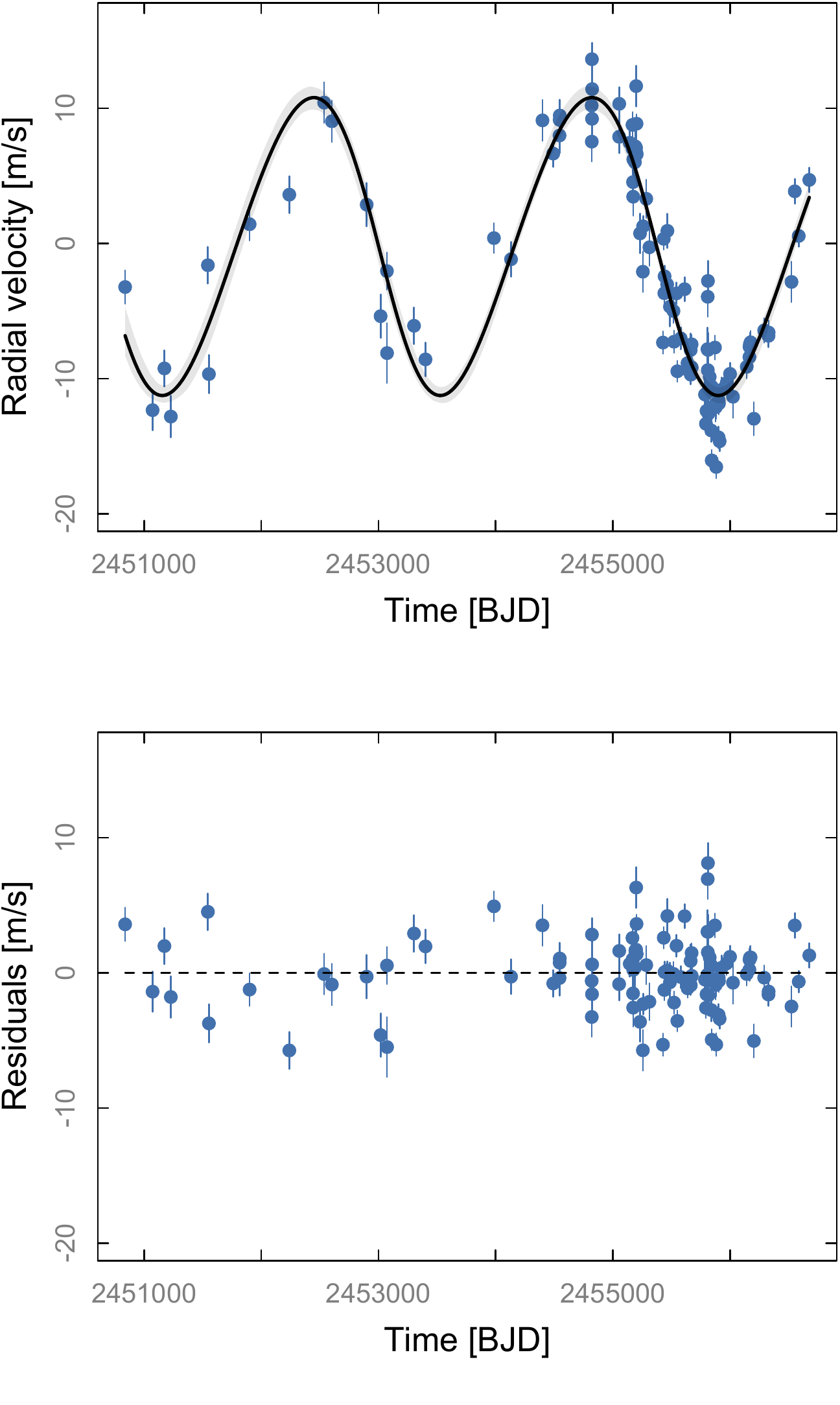}
\plotone{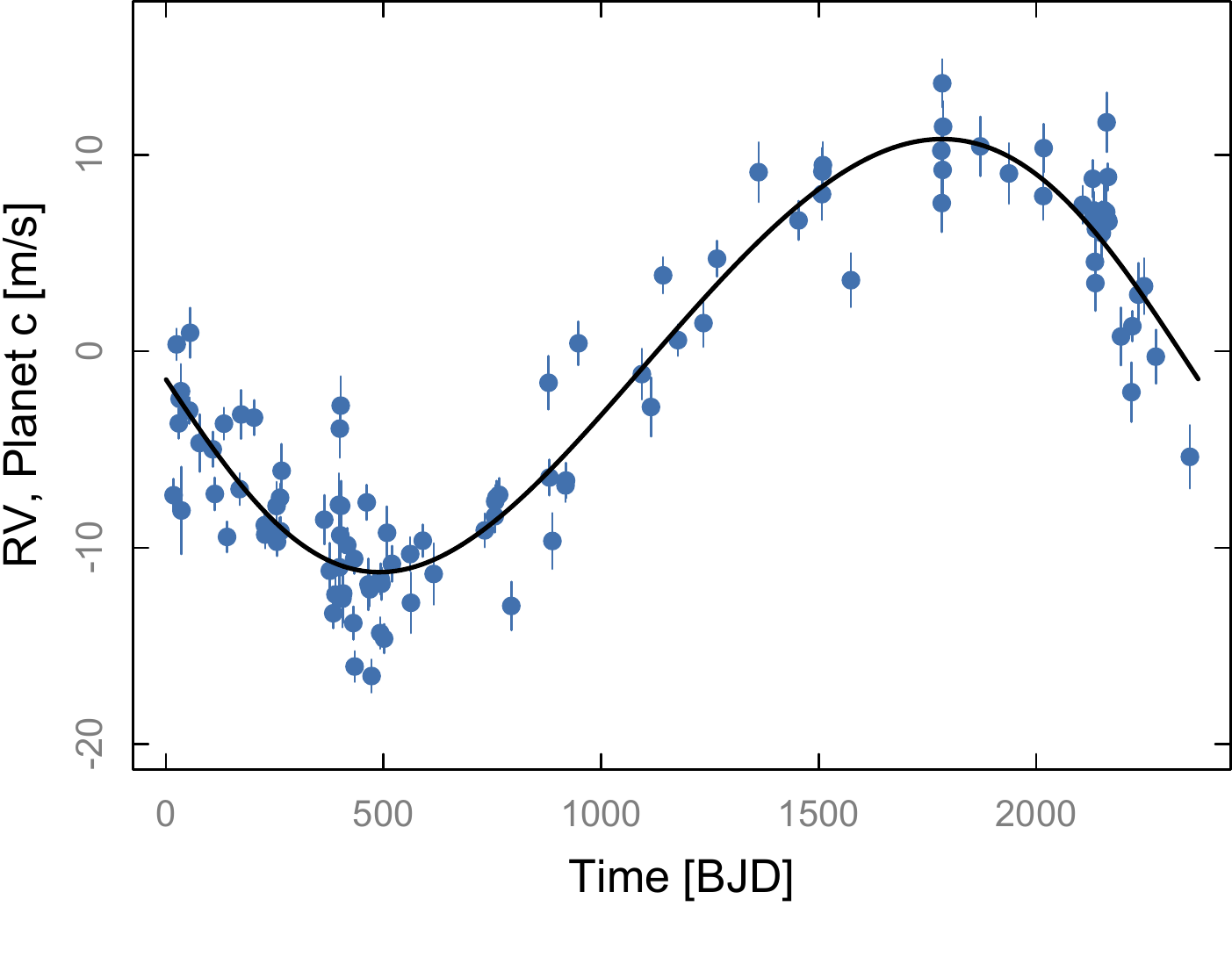}
\plotone{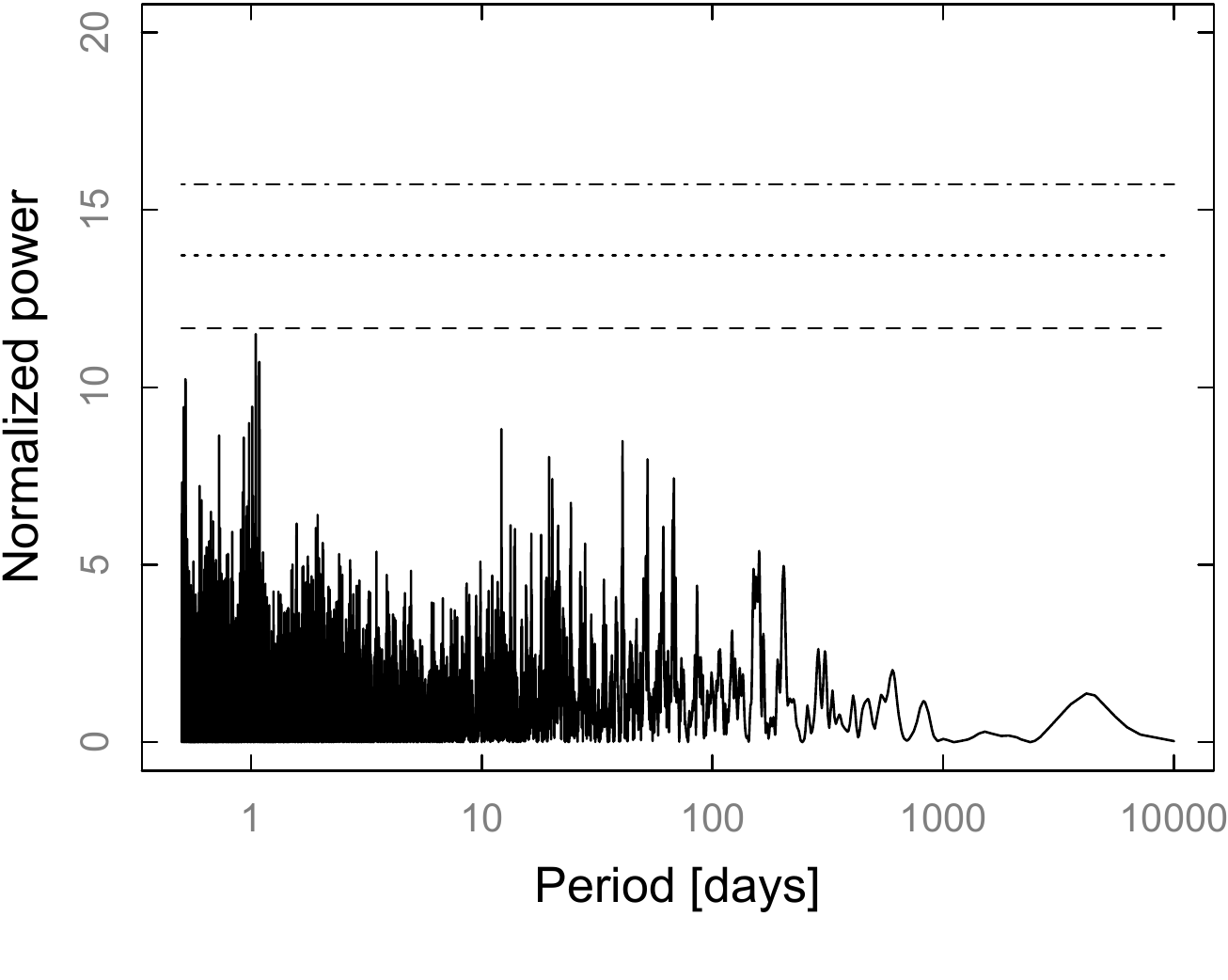}
\caption{\label{fig:bestfit} \panp{First} Best-fit Keplerian model. The gray bands indicate the range of RV response spanned by the model. \panp{Second} RV residuals. \panp{Third} RV observations phased to the best-fit period of planet b. \panp{Fourth} Lomb-Scargle periodogram of the best-fit residuals. }
\end{figure}
\begin{figure}
\epsscale{0.9}
\plotone{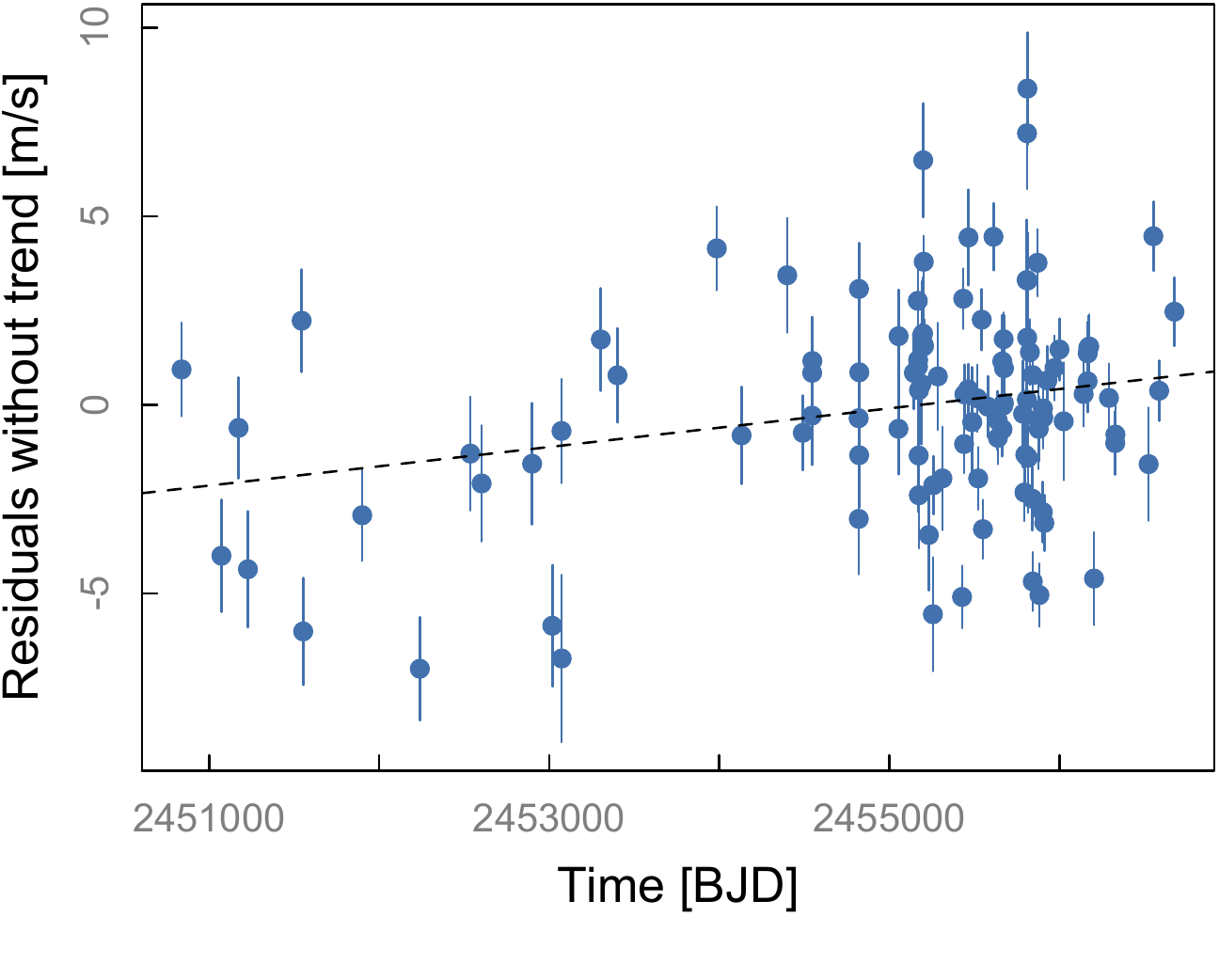}
\caption{\label{fig:trend} RV residuals for model that does not include a linear trend parameter. The best-fit linear trend is superimposed as a dashed line.}
\end{figure}

\begin{figure}
\plotone{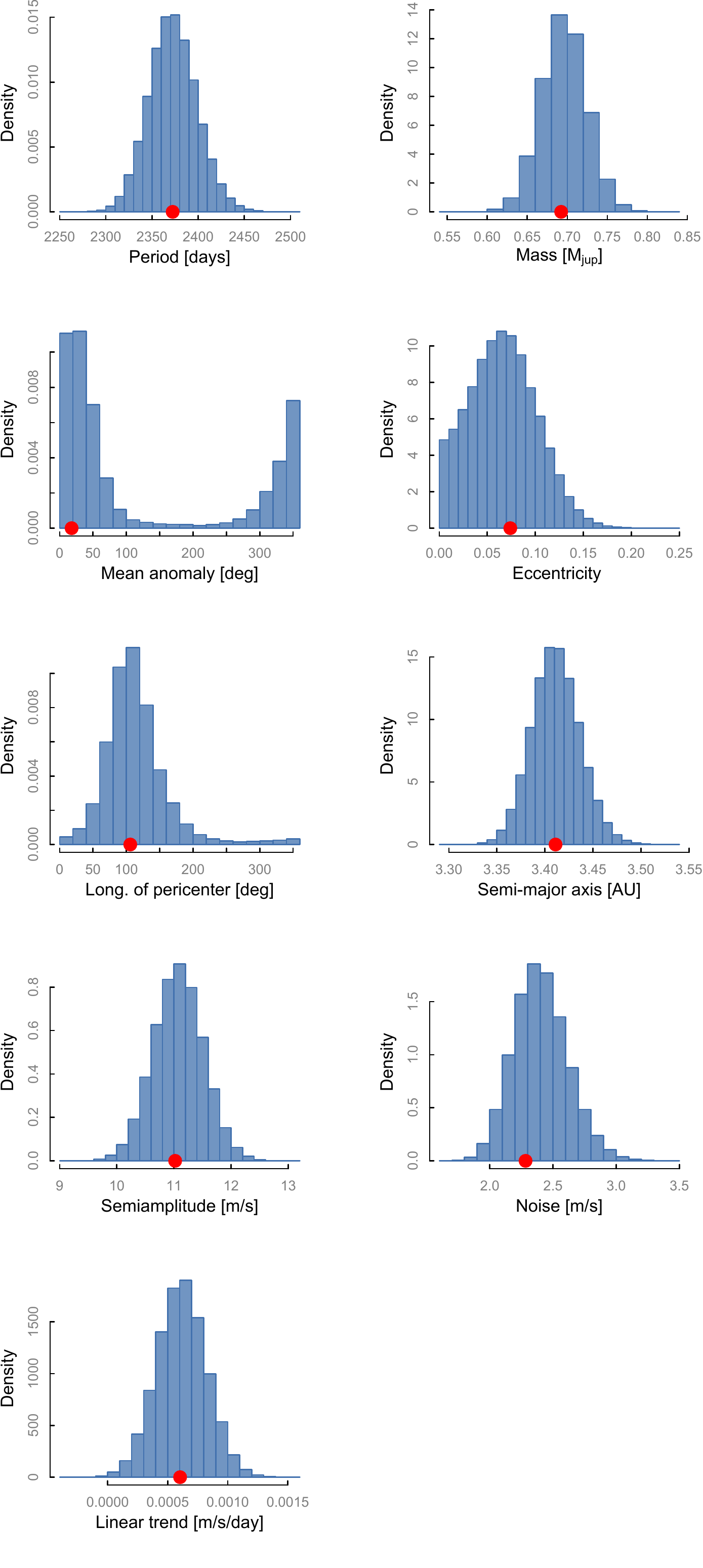}
\caption{\label{fig:dist} Marginal distributions of the orbital elements, drawn from 200,000 Markov-Chain Monte Carlo samples. The red dots mark the set of parameters with the highest likelihood.}
\end{figure}

\begin{figure}
\centering
\begin{framed}
\plotone{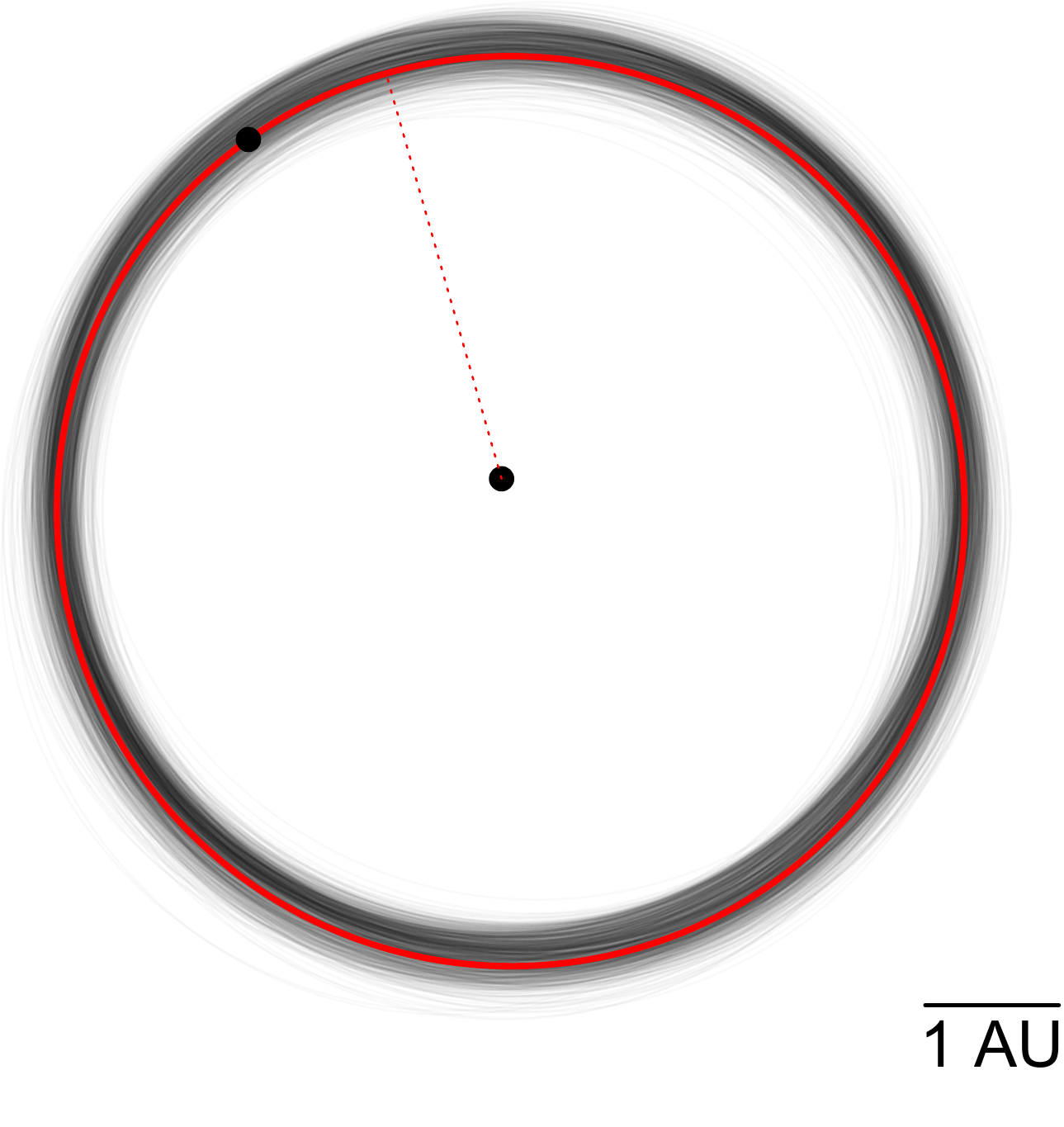}
\end{framed}
\caption{\label{fig:orbit} Orbital plot of the 1-planet model. The best-fit orbit is shown in red; 1,000 samples from the Markov-Chain Monte Carlo run are shown.}
\end{figure}

In order to ascertain the presence of periodic signals in the data, we computed the error-weighted, normalized Lomb-Scargle periodogram \citep{Zechmeister09}, shown in the top panel of Figure \ref{fig:periodograms}. The three horizontal lines in the plot represent different levels of false alarm probability (FAP; 10\%, 1\% and 0.1\%, from bottom to top, respectively). The FAPs were computed by scrambling the dataset 100,000 times and sampling the periodogram at 50,000 frequencies, in order to determine the probability that the power at each frequency could be exceeded by chance \citep[e.g.][]{Marcy05}. The bottom panel of Figure \ref{fig:periodograms} shows the spectral window, displaying the usual peaks due to observational cadence, arising from the sidereal and solar days, the lunar month, and from the solar year \citep{Dawson10}.

The tallest peak at $\approx 2381.04$ days has a FAP $< 2\times 10^{-5}$, much lower than the usual threshold for discovery ($\approx 10^{-3}-10^{-4}$). Our procedure for modeling the signal is the same as that employed in \citet{Vogt15}; here, we briefly repeat the salient parts. We fit the radial velocities with a 1-planet Keplerian model, wherein we vary a vector of parameters consisting of the orbital elements (period, mass, mean anomaly, eccentricity and longitude of pericenter). An additional term $s$ is added in quadrature to the fixed, per-measurement observing errors $e_i$ in order to model additional sources of scatter (e.g. underestimated measurement errors, stellar jitter, and other astrophysical sources of RV variation). The best-fit parameters optimize the log-likelihood of the model:
\begin{equation}
\log \mathcal{L} = -\frac{1}{2} \left[\chi^2 + \sum_{i=1}^{N_{\mathrm{o}}} \log(e_i^2 + s_i^2) + N_{\mathrm{o}} \log(2\pi)\right],\label{eqn:logl}
\end{equation}
where 
\begin{equation}
\chi^2 = \sum_{i=1}^{N_{\mathrm{o}}} {(V_i-v_i)^2}/({e_i^2 + s_i^2})\, ,\label{eqn:chisq}
\end{equation}
and $V_i$ and $v_i$ are the predicted and observed radial velocity measurements at time $t_i$.

A simple Markov-Chain Monte Carlo algorithm (MCMC; e.g. Ford 2005, 2006; Gregory 2011) was used to derive the marginal distribution of the parameters of the model. The output parameters sample the posterior probability density given by Equation \ref{eqn:logl} and flat priors on log $P$, log $\mass$, and the other orbital parameters.

The final best-fit model is shown in Figure \ref{fig:bestfit}. Our model strongly indicates the presence of a single planet candidate (HD~32963b) with $P \approx 2372$ days (6.5 years, $a \approx 3.41$ AU) and $\mass \approx 0.7 \mjup$. The planet's eccentricity is very low ($e \approx 0.07$) and consistent with 0. We note that there is a small linear trend in the residuals ($\approx 0.22 \ms\ \mathrm{year}^{-1}$), which is fitted together with the planet's orbital parameters.

\begin{figure*}
\centering
\plotone{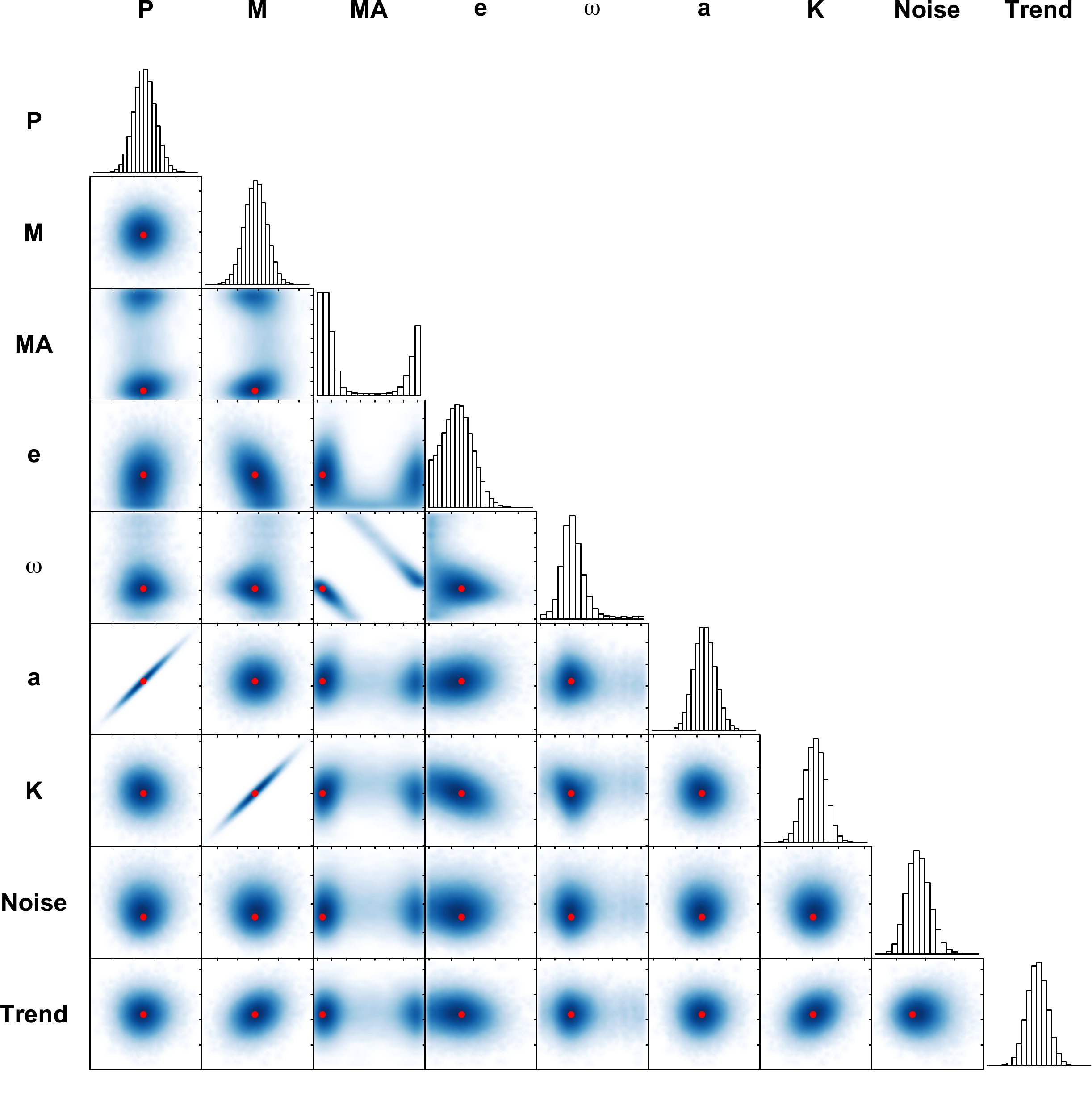}
\caption{\label{fig:corrs} Correlation diagram of the orbital elements, drawn from 200,000 Markov-Chain Monte Carlo samples. The red dots mark the set of parameters with the highest likelihood.}
\end{figure*}

The MCMC algorithm yields a set of fit parameters that sample the posterior distribution. We summarize these distributions by reporting the median and median absolute deviation (MAD) for each parameter in Table \ref{tab:fit}. We also show the marginal distribution of the model parameters in Figure \ref{fig:dist}, and a full pairs correlation plot in Figure \ref{fig:corrs}. 

The bottom panel of Figure \ref{fig:bestfit} shows the Lomb-Scargle periodogram of the residuals for the best-fit model. None of the peaks in the residuals periodogram is significant (FAP $> 0.2$).

\input{HD32963_table}

%% file: HD32963_table.tex
\begin{table}
\centering
\caption{HD~32963 planetary parameters (median and mean absolute deviation)\label{tab:fit}} 
\begin{tabular}{rrl}
  \hline
 & & HD~32963 b \\ 
  \hline
Period [days] & $P$ & 2372 [26] \\ 
  Mass [$\mass_\s{jup}$] & $\mass\sin i$ & 0.70 [0.03] \\ 
  Mean anomaly [deg] & $M$ & 18 [34] \\ 
  Eccentricity  & $e$ & 0.07 [0.04] \\ 
  Longitude of pericenter [deg] & $\varpi$ & 107 [35] \\ 
  Semiamplitude [$\mathrm{m s}^{-1}$] & $K$ & 11.1 [0.4] \\ 
  Semi-major axis [AU] & $a$ & 3.41 [0.02] \\ 
  Periastron passage time [JD] & $T_\s{p}$ & 2450532 [324] \\ 
  Jitter [$\mathrm{m s}^{-1}$] & $s$ & 2.4 [0.2] \\
  Linear trend [$\mathrm{m s}^{-1}\ \mathrm{ year}^{-1}$] & & 0.22 [0.07] \\
  \hline
  Stellar mass [$\mass_\odot$] & & 0.94 \\ 
  RMS [$\mathrm{m s}^{-1}$] & & 2.64 \\ 
  Epoch [JD] & & 2450837.7941 \\ 
  Data points  & & 109 \\ 
  Span of observations [JD] & & 2450837.79 (Jan 1998) \\
                            & & 2456675.91 (Jan 2014) \\ 
   \hline
\end{tabular}
\end{table}

%% file: frequency.tex
\section{The occurrence rate of Jupiter analogs}\label{sec:frequency}
\subsection{A working definition}
The set of RV observations taken with Keck/HIRES, with its relatively uniform biases, long-term monitoring of hundreds of stars, and low formal uncertainties, offers a unique opportunity to probe the occurrence rates of long-period planets. In particular, we are looking to understand the frequency of exoplanets with properties similar to our own Jupiter and HD~32963b.

The first difficulty arises from establishing a definition for a ``Jupiter analog''. This is a difficult and somewhat arbitrary task, since we do not have a full grasp of the population of exoplanets with $P > 10-15$ years. We are interested in choosing orbital parameters such that, conceivably, planets falling in the parameter region chosen will play a similar dynamical role to Jupiter for their planetary system. 

We approximately center our region of interest on Jupiter's parameters. Firstly, we chose to limit periods to the range between 5 and 15 years, between 3 and 6 AU for a solar-mass star ($P_\s{jup} \approx 11.86$ years, $a_\s{jup} \approx 5.2$ AU). The inner limit corresponds to the approximate location of the classical ice line. This range should roughly correspond to planets which formed close to the ice line and did not migrate more than a few AUs. The mass of Saturn (0.3 $\mjup$) provides a natural lower bound for mass. We thus choose to limit our definition of Jupiter analogs to the 0.3-3 $\mjup$ mass range (i.e. within a factor of 3 of Jupiter's mass). Finally, we require the eccentricity to be small ($e < 0.3$), an indication that the system did not undergo a period of strong dynamical instability after the dissipation of the protoplanetary disk.

Table \ref{tab:ja} shows a list of known exoplanets that satisfy the above requirements, as listed by the Exoplanet Data Explorer in August 2015 \citep{Wright11}. The table also contains HD~219134~g \citep{Vogt15} and HD~32963~b. We excluded any star with an unpublished planet from the list.

\input{published_ja}

\subsection{Numerical method}
In this paper, we limit ourselves to a simplistic, yet straightforward re-derivation of occurrence rates for Jupiter analogs, following \citet{Wittenmyer11}.

Our database of RV observations consists of 1,122 stars. In order to establish the number of Jupiter analogs that are detectable using our Keck data alone, we first fit each of the datasets individually, using an automatic fitting procedure analogous to that presented in Section \ref{sec:fit}. This procedure yielded 8 Jupiter analogs (a raw frequency of 0.71\%; HD~134987~c is excluded because of the baseline requirement). 

For each star, we computed synthetic datasets to derive the detectability of planetary signals. We created synthetic planetary signals over a uniform grid in $\log P$, $\log \mass$, $M$, $e$, and $\varpi$ combinations. Looping over the grid generates 320,000 planetary signals, which sample 20 values for $\log P$, $\log \mass$ and $M$, 10 values for $\varpi$, and 4 values for $e$ (0, 0.1, 0.2, 0.3). Each of the 320,000 synthetic datasets is created by computing the Keplerian RV response at the observation times. We then degrade the observations by adding noise. The noise is derived by considering the residuals from the best-fit (a proxy for the residual noise), scrambling them, and adding them to the synthetic RV datasets. We also sample the uncertainty associated with each timestamp to add additional noise to the data.

We then attempt to verify whether the synthetic planet is observable by evaluating the Lomb-Scargle periodograms of each synthetic dataset. We require the peak corresponding to the synthetic planet's period to have a false alarm probability $< 0.1\%$ \citep[computed analytically;][]{Baluev08}, and to match the original period within a 27\% relative tolerance \citep{Wittenmyer11}. We flag a synthetic planet as not detected if the above condition is not satisfied, or if the dataset has less than 10 points, or if the observation baseline is shorter than 1 cycle. Finally, we compute the detectability at each $(\log P, \log\mass, e)$ combination by averaging over the phases. The final output of this procedure is a recovery rate $f_{ijk} = f(P_i, \mass_j, e_k)$ for each star. Each element of $f$ varies between 0 (a planet at the given period, mass, and eccentricity would never be detected from the available data) and 1 (a planet at the given period, mass, and eccentricity would always be detected). 

\begin{figure}
\plotone{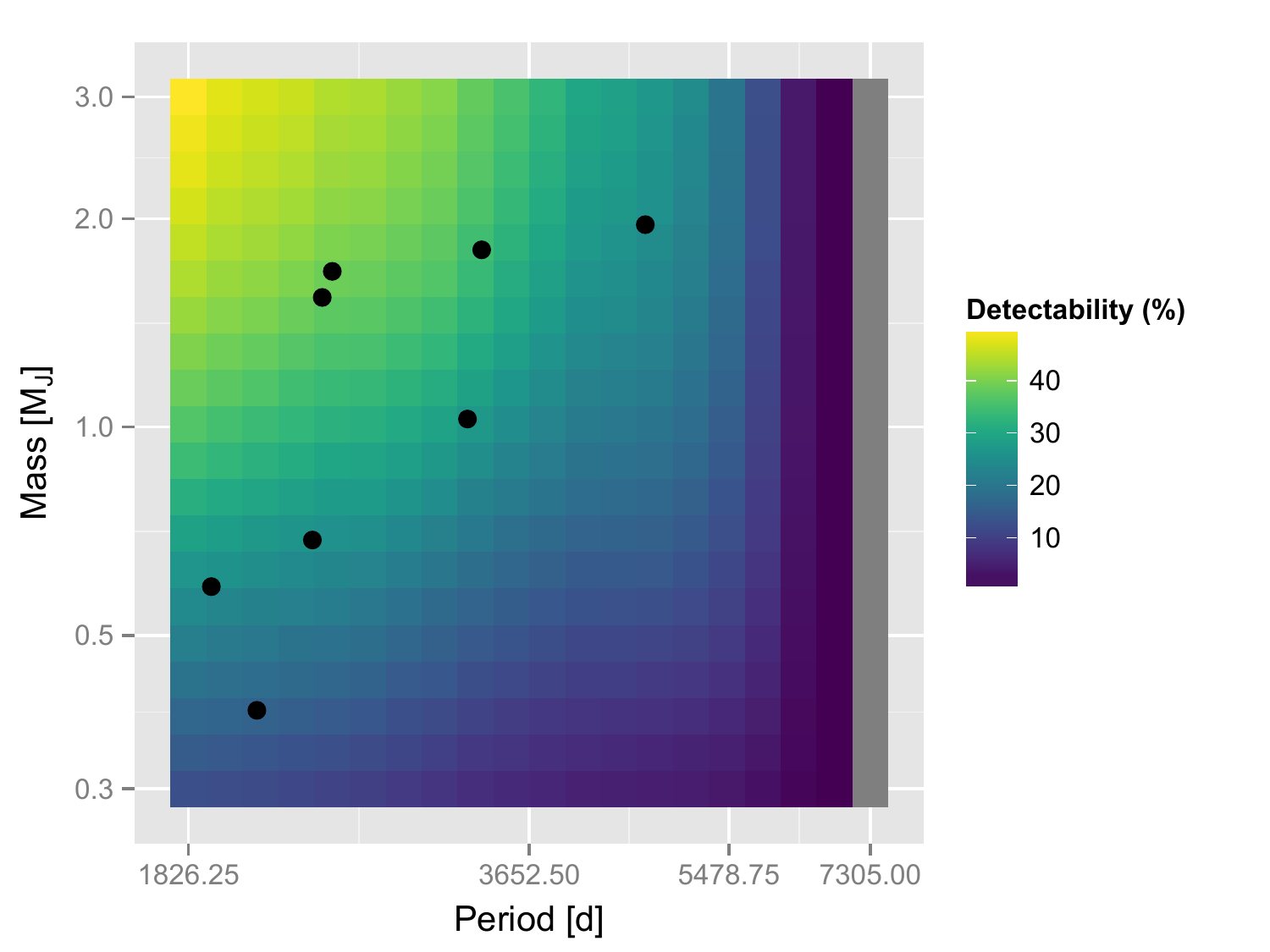}
\caption{\label{fig:freqmatrix} Survey completeness for circular orbits, given by averaging the detection frequencies over all the stars in the Keck dataset at the given period and mass. The grey area indicates periods longer than the baseline of the RV dataset ($P > 18$ years). The black dots mark the Jupiter analogs detected in the Keck sample.}
\end{figure}

Figure \ref{fig:freqmatrix} shows the completeness rate for circular orbits. The completeness is given by 
\begin{equation}
f_\s{C}(P_i, \mass_i, e_i) = \frac{1}{N} \sum_{j=1}^N f_j(P_i, \mass_i, e_i)\ ,
\end{equation}
i.e. the detection frequency averaged over all the stars in our sample. The value for each cell can be interpreted as the fraction of planets that would be detected, if every star in the sample hosted a planet at the cell's period and mass. The median recovery rate is  $\approx 19\%$ over the detectable period range (5-18 years), and $\approx 22\%$ over the period range for Jupiter analogs (5-15 years).

\subsection{Results}
We calculate the approximate frequency of Jupiter analogs for the Keck sample by considering the planets detected, and correcting for the expected completeness at their location in the ($P$, $\mass$, $e$) space. The frequency is given by
\begin{equation}
F_\s{Jup} = \frac{1}{N} \sum_{i=1}^{N_\s{hosts}} \frac{1}{f_i(P_i, \mass_i, e_i) f_\s{C}(P_i, \mass_i, e_i)}\label{eqn:frequency}
\end{equation}
\citep{Wittenmyer11}. In Equation \ref{eqn:frequency}, $f$ is the recovery rate at the given period, mass, and eccentricity of the $i$-th observed Jupiter analog, and $N_\s{hosts}$ and $N$ are the number of stars hosting a Jupiter analog (9) and the total number of stars (1,122), respectively. Summed over the 8 detected Jupiter analogs, we get $F_\s{jup} \approx 3.0\%$.  

In order to compute credible intervals for the occurrence rate, we used an alternative approach based on a simple Monte Carlo calculation. We draw $N_\s{Jup}$ random parameter triplets for Jupiter analogs (comprising period, mass, and eccentricity), assign them to random stars in the sample, observe them with probability given by the recovery rate $f$, and finally require that the output number of observed Jupiter analogs be $N_\s{Jup,\ obs} = 8$. For each input set of size $N_\s{Jup}$, only a fraction will be compatible with the observed number of planets, yielding a broad distribution of $F_\s{Jup}$. This approach has the advantage of yielding credible intervals for the occurrence rate of Jupiter analogs.

\begin{figure}
\plotone{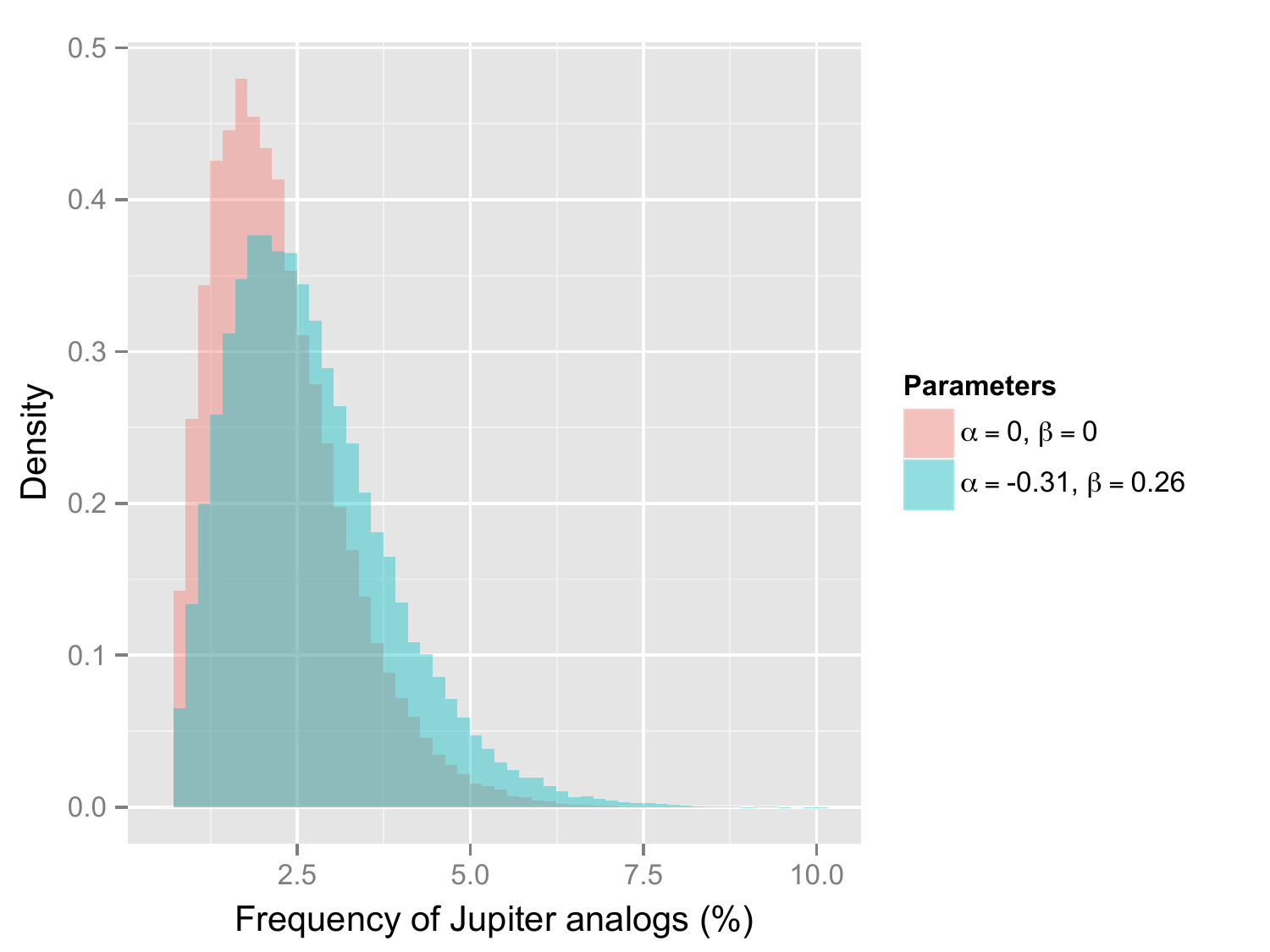}
\caption{\label{fig:freqdist} Occurrence rate distributions for two different sets of $\alpha$ and $\beta$.}
\end{figure}

In order to draw the triplets, we consider a uniform distribution between 0 and 0.3 for eccentricity, and a period and mass distribution given by
\begin{equation}
\frac{dN}{d\log\mass\ d\log P} = N_0 \mass^\alpha P^\beta\ 
\end{equation}
We consider two sets of $\alpha$ and $\beta$: a distribution that is flat in $\log P$ and $\log \mass$ ($\alpha = \beta = 0$), and the values derived by \citet{Cumming08} by fitting the observed distribution ($\alpha = -0.31$, $\beta = 0.26$). The distributions yielded by our approach are shown in Figure \ref{fig:freqdist}, and the 10\%-90\% intervals are reported in Table \ref{tab:ja_freq}. We find that the frequency rate of Jupiter analogs in our sample, based on this calculation, is likely no less than $\approx$ 1\%, and no more than $\approx$ 4\%.

\begin{table}
\centering
\begin{tabular}{rccc}
\hline
 & 10\% & Median & 90\% \\
 \hline
Equation \ref{eqn:frequency} & & 3.3\% & \\
\hline
$\alpha = 0$, $\beta = 0$ & 1.1\% & 1.55\% & 3.5\% \\
$\alpha = -0.31$, $\beta = 0.26$ & 1.3\% & 2.0\% & 4.2\% \\
\hline
\end{tabular}
\caption{Frequency of Jupiter analogs\label{tab:ja_freq}}
\end{table}

%% file: published_ja.tex
\begin{table*}
\centering
\begin{tabular}{lccccl}
  \hline
  Name & Keck? & Period & $\mass \sin i$ & Eccentricity & Reference \\
       &       & [years] & [$\mjup$] & & \\
  \hline
 HD 25171 b &  & 5.05 & 0.96 & 0.08 & \citet{Moutou2011} \\ 
 HD 37124 d &  & 5.10 & 0.69 & 0.16 & \citet{Wright2011} \\ 
 GJ 849 b &  & 5.15 & 0.83 & 0.04 & \citet{Bonfils2013} \\ 
 HD 11964A b & $\checkmark$ &  5.32 & 0.61 & 0.04 & \citet{Wright2009} \\ 
 HD 70642 b &  & 5.66 & 1.91 & 0.03 & \citet{Butler2006} \\ 
 HD 89307 b &  & 5.93 & 1.79 & 0.20 & \citet{Fischer2009} \\
 HD 219134 g & $\checkmark$ & 6.15 & 0.34 & 0.06 & \citet{Vogt15} \\
 HD 32963 b & $\checkmark$ & 6.48 & 0.67 & 0.08 & this work \\
 47 UMa c &  & 6.55 & 0.55 & 0.10 & \citet{Gregory2010} \\ 
 HD 290327 b &  & 6.69 & 2.55 & 0.08 & \citet{Naef2010} \\ 
 HD 50499 b & $\checkmark$ & 6.72 & 1.74 & 0.25 & \citet{Vogt05} \\
 HD 6718 b &  & 6.83 & 1.56 & 0.10 & \citet{Naef2010} \\ 
 HD 117207 b & $\checkmark$ & 7.11 & 1.82 & 0.14 & \citet{Butler2006} \\ 
 HD 154345 b & $\checkmark$ & 9.15 & 0.96 & 0.04 & \citet{Wright2008} \\ 
 GJ 832 b &  & 9.35 & 0.64 & 0.12 & \citet{Bailey2009} \\ 
 HD 154857 c &  & 9.45 & 2.58 & 0.06 & \citet{Wittenmyer2014} \\ 
 HD 187123 c & $\checkmark$ & 10.43 & 1.94 & 0.25 & \citet{Wright09} \\
 HD 222155 b &  & 10.95 & 2.03 & 0.16 & \citet{Boisse2012} \\ 
 mu Ara c &  & 11.51 & 1.89 & 0.10 & \citet{Pepe2007} \\ 
 HD 13931 b & $\checkmark$ & 11.55 & 1.88 & 0.02 & \citet{Howard2010} \\ 
 HD 134987 c & $\checkmark$ & 13.69 & 0.80 & 0.12 & \citet{Jones2010} \\ 
   \hline
\end{tabular}
\caption{List of published Jupiter analogs\label{tab:ja}}
\end{table*}

%% file: HD32963discussion.tex
\section{Discussion}\label{sec:discussion}
The Copernican principle \citep{Bondi52} impresses the expectation that Earth, its position, and the architecture of the Solar System are not unusual, and by extension, it implies that Jupiter analogs are common. Seen from afar, Jupiter is the most detectable planet in the solar system (using either direct imaging, Doppler velocity, or astrometric monitoring). As a consequence, early speculative discussions of planet searches \citep{Black80}, as well as early detection efforts \citep{Campbell88} focused on detecting planets with masses and orbits similar to Jupiter. Observations have shown that giant planets occur within the full range of available parameter space, yet, the Copernican principle continues to strongly influence thinking. For example, there is little direct evidence that long-distance disk migration produces hot Jupiters, but the theory maintains its impetus in part because of the expectation that Jupiter-like planets should form at Jupiter-like distances beyond the protostellar ``ice line'' \citep{Pollack96}.

Notwithstanding the detection of a new Jupiter-like planet on a Jupiter-like orbit around HD~32963, our analysis of over 1000 stars observed at Keck with precision Doppler spectroscopy suggests that Jupiter analogs have an occurrence rate of $\sim$3\% within our sample. Because of the gradual rate of data accrual for long-period candidate planets, this result adds further weight to a consensus that has been emerging for some time. \citet{Cumming08} analyzed 585 stars in the Keck Planet Search with $N>10$ velocities (representing an 8-year subset of the data used in this study), and found an occurrence rate of less than 10.5\% for planets with $0.3\,\mjup<\mass_{\rm P}<10\,\mjup$ and $P<2000$ days; when extrapolated out to $20$ AU, they also estimated an occurrence rate of $\approx 3\%$. \citet{Gould10}, using the results of microlensing surveys, estimated a frequency $f\sim1/6$ of systems orbiting $M_{\star}\sim0.5M_{\odot}$ stars that resemble the outer solar system (in that they contain planets with ice-giant masses and above in the region $a\sim3.5$AU. \citet{Wittenmyer11}, working with data obtained at the Anglo-Australian Telescope, found that approximately 3\%, and  no more than 37\%, of stars host a giant planet between 3 and 6 AU from the parent star. Our result is consistent with all of these studies.

The relative paucity of Jovian planets with orbital periods of several thousand days has several seemingly straightforward implications. First, it could indicate that the migration of giant planets from beyond the ice line of the protoplanetary disk is indeed efficient, and that whenever Jupiter-mass planets form at $a\sim5$ AU, they have little chance of staying there. In this view, the Grand Tack scenario \citep{Walsh11} provides an explanation for Jupiter's current location by positing that the establishment of a 3:2 mean-motion resonance with Saturn fortuitously compelled Jupiter to return to the vicinity of its formation zone. Alternately, it might be difficult to fully complete core accretion at $a\sim5$ AU in a typical protostellar disk. Accretion calculations \citep[e.g.][]{Hubickyj05} suggest that proto-giant planet cores require disks with a high surface density of solids in order to form quickly enough to initiate rapid gas accretion within the confines of a typical disk lifetime. High-metallicity disks have higher surface densities of solids, and are presumably associated with high metallicity stars. This correlation is likely the source of the giant-planet stellar metallicity connection \citep{Gonzalez08, Santos01, Fischer05}, and suggests that true Jupiter analogs might be common only among stars within a particular metallicity range.

Long-running radial velocity programs such as the Keck survey will continue to play a vital role in determining the architectures of the outer planetary systems orbiting nearby solar-type stars. Our work indicates that substantial efforts are still required to see a majority of Jupiter-like planets, despite the seemingly impressive $K=12\,{\rm m s^{-1}}$ radial velocity half-amplitude that Jupiter induces in the Sun). Substantially more concerted efforts will be required to probe the frequency and orbital properties of ice giants with $P>2000$ days. The planet Neptune, for example, if placed at 5 AU, would induce a mere $64\,{\rm cm\,s^{-1}}$ signal. While potentially feasible, detections of such planets will require an excellent understanding of all sources of systematic error, both ground-based and astrophysical.

%% file: acknowledgments.tex
\section{Acknowledgments}
We would like to thank Dr. Ramirez for useful discussions, and our referee for their useful critique.
 
DR's research was conducted under the Byram Hills Authentic Science Research Program.
SM acknowledges support from the W. J. McDonald Postdoctoral Fellowship and the Longhorn Innovation Fund for Technology grant. This material is based upon work supported by the National Aeronautics and Space Administration through the NASA Astrobiology Institute under Cooperative Agreement Notice NNH13ZDA017C issued through the Science Mission Directorate. GL acknowledges support from the NASA Astrobiology Institute through a cooperative agreement between NASA Ames Research Center and the University of California at Santa Cruz, and from the NASA TESS Mission through a cooperative agreement between M.I.T. and UCSC. SSV gratefully acknowledges support from NSF grants AST-0307493 and AST-0908870. RPB gratefully acknowledges support from NASA OSS Grant NNX07AR40G, the NASA Keck PI program, and from the Carnegie Institution of Washington. The work herein is based on observations obtained at the W. M. Keck Observatory, which is operated jointly by the University of California and the California Institute of Technology, and we thank the UC-Keck and NASA-Keck Time Assignment Committees for their support. This research has additionally made use of the Keck Observatory Archive (KOA), which is operated by the W. M. Keck Observatory and the NASA Exoplanet Science Institute (NExScI), under contract with the National Aeronautics and Space Administration. We wish to extend our special thanks to those of Hawaiian ancestry on whose sacred mountain of Mauna Kea we are privileged to be guests. Without their generous hospitality, the Keck observations presented herein would not have been possible. The authors acknowledge the Texas Advanced Computing Center (TACC, \url{http://www.tacc.utexas.edu}) at The University of Texas at Austin for providing HPC resources that have contributed to the research results reported within this paper. This research has made use of the SIMBAD database, operated at CDS, Strasbourg, France, and the Exoplanet Orbit Database and the Exoplanet Data Explorer at exoplanets.org. This paper was produced using $^BA^M$.